\documentclass[nofootinbib,pra,twocolumn,superscriptaddress]{revtex4-2}
\usepackage{float,graphicx,multirow, amsmath,color,dsfont}
\usepackage{amsmath,bm}
\usepackage{enumerate}
\usepackage{amssymb,mathrsfs,dsfont}
\usepackage{amssymb,mathbbol}
\usepackage{amsfonts}
\usepackage{mathrsfs}
\usepackage{float}
\usepackage{xcolor,hyperref}
\definecolor{darkblue}{rgb}{0,0.0.1,0.3}
\definecolor{darkred}{rgb}{0.6,0.1,0}
\hypersetup{colorlinks,breaklinks,
	linkcolor=darkred,urlcolor=darkblue,
	anchorcolor=darkred,citecolor=
	darkred,pdfauthor=ChRiSh, pdftitle=ChRiMoSh.SVS.Estimation}
\usepackage{tikz}
\usepackage{mathbbol}
\usepackage{float}
\usepackage[normalem]{ulem} 

\newcommand{\ie}{\textit{i}.\textit{e}.}

\begin{document}
	
	\title{Parity-detection-based Mach-Zehnder interferometry with 
coherent and non-Gaussian squeezed vacuum states as inputs}
	\author{Chandan Kumar}
	\email{chandan.quantum@gmail.com}
	\affiliation{Department of Physical Sciences,
		Indian
		Institute of Science Education and
		Research Mohali, Sector 81 SAS Nagar,
		Punjab 140306 India.}
	\author{Rishabh}
	\email{rishabh1@ucalgary.ca}
	\affiliation{Department of Physics and
		Astronomy, University of Calgary, Calgary T2N1N4, Alberta,
		Canada.}
			\author{Mohak Sharma}
	\email{mohak.quantum@gmail.com}
	\affiliation{Department of Physical Sciences,
		Indian
		Institute of Science Education and
		Research Mohali, Sector 81 SAS Nagar,
		Punjab 140306 India.}	
	\author{Shikhar Arora}
	\email{shikhar.quantum@gmail.com}
	\affiliation{Department of Physical Sciences,
		Indian
		Institute of Science Education and
		Research Mohali, Sector 81 SAS Nagar,
		Punjab 140306 India.}	
	
	\begin{abstract}
We theoretically explore the advantages rendered by non-Gaussian operations in phase estimation using a parity-detection-based Mach-Zehnder interferometer, with one input being a coherent state and the other being a non-Gaussian squeezed vacuum state (SVS). We consider a realistic model to perform three different non-Gaussian operations, namely photon subtraction, photon addition, and photon catalysis on a single-mode SVS. We start by deriving the Wigner function of the non-Gaussian SVSs, which is then utilized to derive the expression for the phase sensitivity. The analysis of the phase sensitivity reveals that all three different non-Gaussian operations can enhance the phase sensitivity under suitable choices of parameters. We also consider the probabilistic nature of these non-Gaussian operations, the results of which reveal the single photon addition to be the optimal operation. Further, our analysis also enables us to identify the optimal squeezing of the SVS and the transmissivity of the beam splitter involved in the implementation of the non-Gaussian operations.

	\end{abstract}
	\maketitle

	\section{Introduction}

 Mach-Zehnder interferometer (MZI) is the most commonly employed optical instrument in phase measurement~\cite{Dowling-cp-2008, Giovannetti2011}. If the input beams to the MZI are classical sources, the phase sensitivity is bounded by shot-noise limit (SNL)~\cite{caves-prd-1981}.    To improve the phase sensitivity, quantum resources such as N00N states, twin Fock states, and squeezed states have been employed. These quantum resources enable the phase sensitivity to go beyond   SNL and reach the Heisenberg limit (HL)~\cite{Dowling,Giovannetti-science-2004,Hofmann-pra-2007, Anisimov-prl-2010,caves-prl-2013,Jeong-prl-2019}.   
 
 The maximum squeezing that can be achieved experimentally is bounded~\cite{15dB}, which leads to a limited enhancement in the phase sensitivity. To overcome this drawback, one can resort to non-Gaussian (NG) operations such as photon subtraction (PS), photon addition (PA), and photon catalysis (PC).  It has already been shown that NG operations can be beneficial in quantum teleportation~\cite{tel2000,Akira-pra-2006,Anno-2007,tel2009,wang2015,catalysis15,catalysis17,tele-arxiv},  quantum key distribution~\cite{qkd-pra-2013,qkd-pra-2018,qkd-pra-2019,qk2019, zubairy-pra-2020}, quantum illumination~\cite{illumination14}, and quantum metrology~\cite{gerryc-pra-2012,josab-2012,braun-pra-2014,josab-2016,pra-catalysis-2021,crs-ngtmsv-met,metro-thermal-arxiv}.
 
 In particular, Ref.~\cite{gerry14} showed that the phase sensitivity of parity-detection-based MZI at a fixed squeezing could be enhanced when the inputs are coherent state and ideal photon-subtracted SVS as compared to the case when coherent state and SVS are employed as the inputs.

 In this article, we extend the analysis of~\cite{gerry14} to a wider class of NG states. To generate these NG states, we perform three distinct NG operations namely PS, PA, and PC  on SVS. We implement these  NG operations via a realistic model based on multiphoton Fock states, photon number resolving detectors, and beam splitter [Fig.~\ref{mzi}].  
 This leads to the generation  of three distinct families of states namely  photon-subtracted SVSs (PSSVSs), photon-added SVSs (PASVSs), and photon-catalyzed SVSs (PCSVSs), which we collectively term as ``NGSVSs".

 We then evaluate the Wigner function of these NGSVSs,   where the free parameters   include    input Fock state, detected Fock state, and the transmissivity of the beam splitter involved in the implementation of the NG operation.
 By suitably choosing the input Fock state and detected Fock state, we can perform either PS, PA, or PC operation on SVS. 
 The Wigner function is then utilized to evaluate the expression of the phase sensitivity  for parity-detection-based MZI.  
 
 We analyze the behavior of the phase sensitivity of NGSVSs as a function of different parameters. 
 The analysis  reveals that all three NG operations can lead to a significant enhancement under suitable choices of parameters.
 Further, we take the probabilistic nature of NG operations into account in our analysis, which reveals that single photon-added SVS is the optimal state.

 It should also be noted that the PS operation considered in Ref.~\cite{gerry14} is implemented by annihilation operator $\hat{a}$, which is nonphysical. In contrast, our realistic scheme for the implementation of NG operations  can be realized with current technologies, including multiphoton Fock state~\cite{singlephoton,single-photon,singlephot,2phton,3phton} and photon number resolving detectors~\cite{Lita:08,Marsili2013,Zadeh}. We would like to point out that this realistic model invariably enhances the complexity of our calculation. Further, the phase sensitivity expression derived here is quite general and special cases investigated in Refs.~\cite{gerry14,photonadded2019} can be obtained in the appropriate limit. Furthermore, the realistic scheme enables us to consider the probabilistic nature of the involved NG operations.

 The rest of the paper is structured as follows. In Sec.~\ref{sec:mzi}, we derive the phase sensitivity expression for the 
 parity-measurement-based MZI with coherent state and NGSVSs as the two inputs. 
 Sec.~\ref{results} carries out the analysis of  the phase sensitivity  to find out the optimal NG operation. 
 We outline our main results and provide directions for future research in  Sec.~\ref{sec:conc}.
 In the Appendix~\ref{app:wigner}, we have provided a detailed calculation of the Wigner distribution function of NGSVSs.

 \section{Parity-measurement-based phase estimation}\label{sec:mzi}

 \begin{figure}[h!]
 	\begin{center}
 		\includegraphics[scale=1]{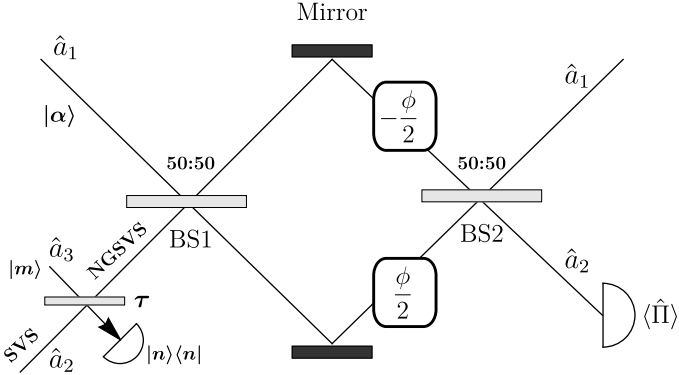}
 		\caption{  Schematic diagram for the implementation of NG operations on a SVS followed by   parity-detection-based  MZI. 
 		The SVS and the ancilla Fock state $|m\rangle$  are  combined using a beam splitter of transmissivity $\tau$, and subsequently, detection of $n$ photons in the ancilla output mode heralds the generation of  NGSVSs. A coherent state and the NGSVSs serve as the resource states of the MZI for the estimation of the introduced phases.
 		}
 		\label{mzi}
 	\end{center}
 \end{figure}
 
 Consider the setup of a lossless MZI shown in Fig.~\ref{mzi}, which consists of two $50{:}50$ beam splitters and two phase shifters. While one of the input states is a coherent state, the other input state is generated by performing different NG operations on a SVS, as depicted in the lower left-hand corner of Fig.~\ref{mzi}.
 An unknown phase $\phi$ is introduced via the two phase shifters, and we aim to estimate this unknown phase by parity detection on the output mode $\hat{a}_2$.
 The Wigner distribution function of the coherent state $| \alpha \rangle$ can be written as
\begin{equation}\label{coh}
    W_{| \alpha \rangle} (\xi_1 ) = (\pi)^{-1}\exp \left[ -(q_1-d_x)^2-(p_1-d_p)^2\right],
\end{equation} 
where $\xi = (q_1,p_1)^T$, and $\alpha=(d_x+id_p)/\sqrt{2}$.
To implement the NG operations, we mix the SVS and the ancilla Fock state $|m\rangle$ via a beam splitter of transmissivity $\tau$. A photon number resolving detector is used to perform a conditional measurement  of $n$ photons   on the ancilla output mode, which signals the generation of NGSVSs. 

For convenience,  we employ phase space formalism, specifically the Wigner distribution function, for calculations. While step-wise calculation for the derivation of the Wigner distribution function of the NGSVSs is provided in the Appendix~\ref{app:wigner}, here we provide  the final expression.  The  Wigner distribution function of the NGSVSs turns out to be (given in Eq.~(\ref{app:normPS}) of the Appendix~\ref{app:wigner})
 \begin{equation}\label{normPS}
	W^{\text{NG}}(\xi_2) = \frac{\bm{\widehat{F}_1} \exp \left(\dfrac{w_1^2 q_2 + w_2^2 p_2 + \bm{u}^T M_1 \bm{u}+ \bm{u}^T M_2}{-w_1 w_2}\right)}{P^{\text{NG}} \sqrt{w_1 w_2} } ,
\end{equation}
where $  w_{1,2} = \cosh{r} \pm \tau \sinh{r}$ and $\bm{u}=(u_1,v_1,u_2,v_2)^T$ represents column vector. Further,  
\begin{equation}
	\bm{\widehat{F}_1} = \frac{(-2)^{m+n}}{\pi \, m! \, n!} \frac{\partial^{m}}{\partial\,u_1^{m}} \frac{\partial^{m}}{\partial\,v_1^{m}} \frac{\partial^{n}}{\partial\,u_2^{n}} \frac{\partial^{n}}{\partial\,v_2^{n}}\{ \bullet \}_{\substack{u_1= v_1=0\\ u_2= v_2=0}},\\
\end{equation}represents differential operator. The explicit form of the matrices $M_1$ and $M_2$ is provided in Eqs.~(\ref{wm1}) and~(\ref{wm2})  of Appendix~\ref{app:wigner}. 
 The success probability of the NG operations, $P^{\text{NG}}$,   is given by (Eq.~(\ref{probeqq}) of the Appendix~\ref{app:wigner})
\begin{equation}\label{probq}
	P^{\text{NG}}= \int d^2 \xi_2  \widetilde{W}^{\text{NG}}_{A' }
	=\frac{ \pi \, \bm{\widehat{F}_1}}{\sqrt{w_1 w_2}} \exp \left(\frac{\bm{u}^T M_3 \bm{u}}{-4 w_1 w_2}\right),
\end{equation} where the matrix   $M_3$ is given in Eq.~(\ref{wm3}) of Appendix~\ref{app:wigner}. 
Different NG operations on SVS can be implemented by fixing the   input Fock state and detected  number of photons. We can perform PS, PA or PC operation on SVS under the condition $m<n$, $m>n$ or $m=n$, respectively. In this article, we set $m=0$ and $n=0$ for PS and PA operations, respectively. These NG operations convert the SVS state from Gaussian to non-Gaussian. 
   
   The derived Wigner distribution function of the SVS~(\ref{normPS}) is quite general and   Wigner distribution function of special states can be obtained in different limits. For example, the Wigner distribution function of the ideal PSSVSs can be obtained in the limit $\tau \rightarrow 1$ with  $m=0$. The ideal PSSVSs are represented by $ \mathcal{N}_s \hat{a}^{n}  |\text{SVS}\rangle$, where $\mathcal{N}_s$ is the normalization factor.
	Similarly, 	the Wigner distribution function of the ideal PASVSs can be obtained in the limit $\tau \rightarrow 1$ with  $n=0$. 
		The ideal PASVSs are represented by $ \mathcal{N}_a \hat{a}{_2^{\dagger }}^{m}  |\text{SVS}\rangle$, where $\mathcal{N}_a$ is the normalization factor.

For the purpose of ease in the description of the collective action of the MZI, we consider the Schwinger representation of $\text{SU}(2)$ algebra~\cite{yurke-1986}. 
In terms of the annihilation and creation operators of the input modes, the generators of the $\text{SU}(2)$ algebra turn out to be 
\begin{equation}
	\begin{aligned}
		\hat{J}_1 = &\frac{1}{2}(\hat{a}^\dagger_1\hat{a}_2+\hat{a}_1\hat{a}^\dagger_2),\\
		\hat{J}_2 = &\frac{1}{2i}(\hat{a}^\dagger_1\hat{a}_2-\hat{a}_1\hat{a}^\dagger_2),\\
		\hat{J}_3 = &\frac{1}{2}(\hat{a}^\dagger_1\hat{a}_1-\hat{a}^\dagger_2\hat{a}_2).
	\end{aligned}
\end{equation}
These generators are
 also known as angular momentum operators and satisfy
 the commutation relations $[J_i,J_j] = i \epsilon_{ijk}J_k $.  The unitary operators acting on the Hilbert space corresponding to the first  and the second  beam splitters are given by $e^{-i(\pi/2)\hat{J}_1}$ and $e^{i(\pi/2)\hat{J}_1}$, respectively. The combined action of the two phase shifters is represented by the unitary operator $e^{i \phi \hat{J}_3}$. Therefore, the total action of the MZI is represented as a product of   the unitary operators as follows:  
\begin{equation}
	\mathcal{U}(S_{\text{MZI}}) = e^{-i(\pi/2)J_1}e^{i \phi J_3}e^{i(\pi/2)J_1}=e^{-i \phi J_2}.
\end{equation}
The corresponding symplectic matrix $S_{\text{MZI}}$   transforming the phase space variables $(\xi_1,\xi_2)^T$ turns out to be
\begin{equation} 
	S_{\text{MZI}} =  \begin{pmatrix}
		\cos (\phi/2) \,\mathbb{1}& -	\sin (\phi/2) \,\mathbb{1} \\
		\sin (\phi/2) \,\mathbb{1}& \cos (\phi/2) \,\mathbb{1}
	\end{pmatrix}.
\end{equation}
The evolution of the Wigner distribution function due to $S_{\text{MZI}}$ can be stated as~\cite{arvind1995, weedbrook-rmp-2012}
 \begin{equation}
	W_{\text{in}}(\xi ) \rightarrow W_{\text{in}}(S_{\text{MZI}}^{-1}\xi) =W_{\text{out} } (\xi),
\end{equation}
where $W_{\text{in}}(\xi ) = W_{| \alpha \rangle} (\xi_1 )\times W^{\text{NG}}(\xi_2 ) $ is the product of the Wigner distribution function of the coherent state~(\ref{coh}) and NGSVSs~(\ref{normPS}). 
  We employ parity detection on the output mode $\hat{a}_2$ as depicted in Fig.~\ref{mzi}. 
  The operator corresponding to parity detection is given by  
\begin{equation}
	\hat{\Pi}_{\hat{a}_2} =  \exp\left( i \pi   \hat{a}^{\dagger}_2 \hat{a}_2 \right)=  (-1)^{\hat{a}_2^{\dagger}\hat{a}_2}.
\end{equation}
To evaluate the average of the parity operator, we recall that the Wigner distribution function can be expressed as the average of the displaced parity operator~\cite{parity-1977}:
	\begin{equation}\label{wigparity}
		W(\bm{\xi}) =\frac{1}{{ \pi}^{n}} \text{Tr} \left[ \hat{\rho}\, D(\bm{\xi}) \hat{\Pi} D^{\dagger} (\bm{\xi}) \right] ,
	\end{equation}
	where $n$ is the number of modes, $D(\bm{\xi}) = \exp[i \hat{ \xi} \, \Omega \, \bm{\xi}]$ represents the displacement operator, and $ \hat{\Pi} =\prod_{i=0}^{n}  \exp\left( i \pi   \hat{a}^{\dagger}_i \hat{a}_i \right)$
	represents the parity operator. Hence, the average of the parity operator in terms of the Wigner distribution function turns out to be~\cite{Birrittella-2021}:
\begin{equation}\label{apari1}
	\langle \hat{\Pi}_{\hat{a}_2} \rangle   = \pi \int \, d^2\xi_1 \, W_{\text{out}} (\xi_1,0).
\end{equation}
Equation~(\ref{apari1}) evaluates to
\begin{equation}\label{apari}       
	\langle \hat{\Pi}_{\hat{a}_2} \rangle = \frac{ \pi \, \bm{\widehat{F}_1}}{\sqrt{w_3 w_4}} \exp \left(\frac{\bm{u}^T M_4 \bm{u}+\bm{u}^T M_5 \bm{d}+\bm{d}^T M_6 \bm{d}}{-w_3 w_4}\right), 
\end{equation}
where $w_{3,4} = \cosh{r} \pm \tau \sinh{r} \cos{\phi}$ and   $\bm{d}=(2\,d_x,2\,d_p)^T$. Further, the matrices $M_4$, $M_5$, and $M_6$ are defined in the Eqs.~(\ref{wm4}),~(\ref{wm5}) and~(\ref{wm6}) of Appendix~\ref{appsec}.
 The phase uncertainty or sensitivity  can be expressed as following using the error propagation formula:
\begin{equation}\label{phasesens}
	\Delta \phi = \frac{\sqrt{1-	\langle \hat{\Pi}_{\hat{a}_2} \rangle ^2}}{|\partial  	\langle \hat{\Pi}_{\hat{a}_2} \rangle/\partial \phi|}.
\end{equation}
The phase uncertainty  is a function of the squeezing $r$ of the SVS, displacement   $ d_x $ and $ d_p $ of the coherent state, and introduced unknown phase $\phi$. Besides,  the number of input photons $m$ and the number of detected photons $n$ can be appropriately chosen to perform different NG operations. One important advantage of our considered realistic model for the implementation of NG operations is that it allows us to consider the probability of different NG operations and consequently identify their effectiveness in phase estimation.
 
   In the   unit transmissivity limit ($\tau \rightarrow 1$) and $m=0$, the phase sensitivity expression~(\ref{phasesens}) reduces to that of   ideal PSSVSs~\cite{gerry14}.  Similarly,  in the unit transmissivity limit and $n=0$, we obtain the phase sensitivity expression for ideal PASVSs~\cite{photonadded2019}.

	 \section{Phase sensitivity enhancement via NGSVSs}\label{results}

\begin{figure}
	\begin{center}
		\includegraphics[scale=1]{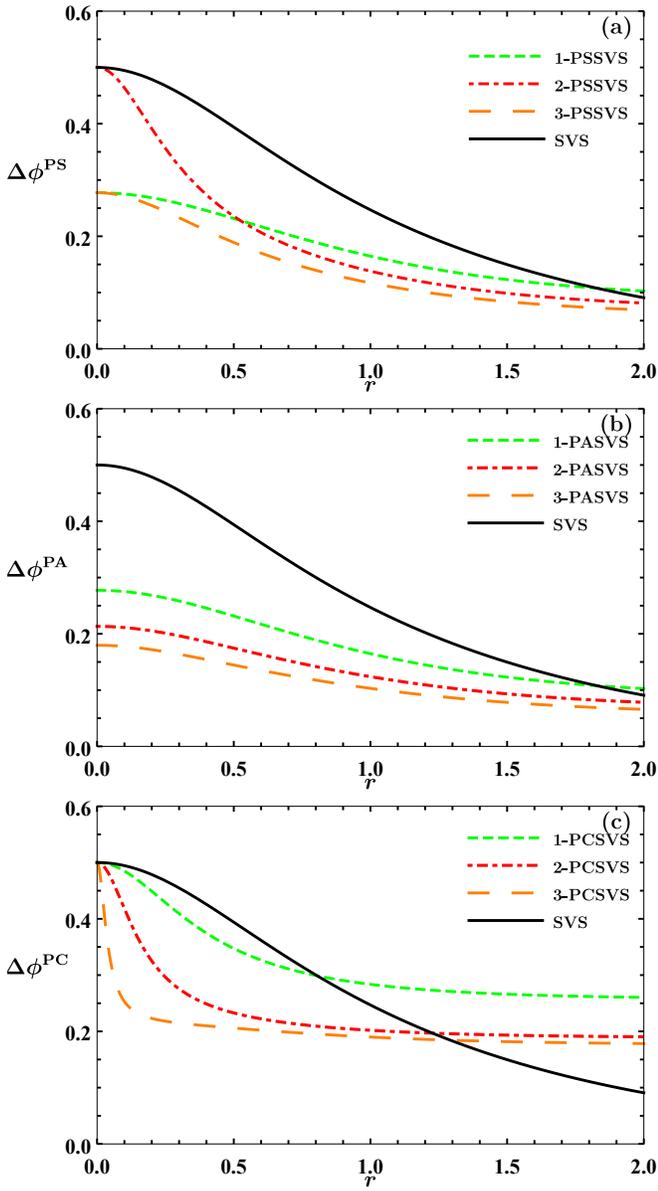}
		\caption{Phase uncertainty $\Delta \phi $ as a function of the  squeezing parameter $r$ for NGSVSs. We have set the transmissivity of the beam splitter to be $\tau=0.9$ for panels (a) and (b) and $\tau=0.1$  for panel (c). Further, the  coherent state displacement has been taken to be $d_x=d_p=2$ and phase  $\phi=0.01$ for all the panels.    }
		\label{phase_1d_sq}
	\end{center}
\end{figure}

	 We now proceed to find out whether different NG operations on SVS can enhance phase sensitivity in MZI. To   this end, we study the behavior of phase uncertainty, $\Delta \phi $, as a function of   initial squeezing ($r$) of SVS, transmissivity ($\tau$)  of beam splitter used to perform NG operations, and   magnitude of the total unknown phase   ($\phi$) introduced in the interferometer. In Fig.~\ref{phase_1d_sq}, we show the plot of $\Delta \phi $ as a function of squeezing, while other parameters are kept fixed\footnote{In this paper, we set the displacement of the coherent state $d_x=d_p=2$ for numerical analysis purposes.}.
As can be seen in Fig.~\ref{phase_1d_sq}(a), PSSVSs improve the phase sensitivity as compared to SVS for almost the complete  range of considered squeezing range. For 1-PSSVS, phase sensitivity improvement is not observed for $r \gtrapprox 1.8$. The phase sensitivity via 2-PSSVS gets better than 1-PSSVS at a certain threshold squeezing. The phase sensitivity of 3-PSSVS is better than 1-PSSVS and 2-PSSVS.

We observe from Fig.~\ref{phase_1d_sq}(b) that 1-PASVS  significantly improves the phase sensitivity up to the squeezing value of $r \approx 1.8$. The phase sensitivity enhances  further as more photons are added. Similarly, the PCSVSs yield better phase sensitivity as we catalyze more photons. However, the phase sensitivity is improved compared to the initial SVS for a much smaller range of the squeezing parameter, as shown in Fig.~\ref{phase_1d_sq}(c).

\begin{figure}
	\begin{center}
		\includegraphics[scale=1]{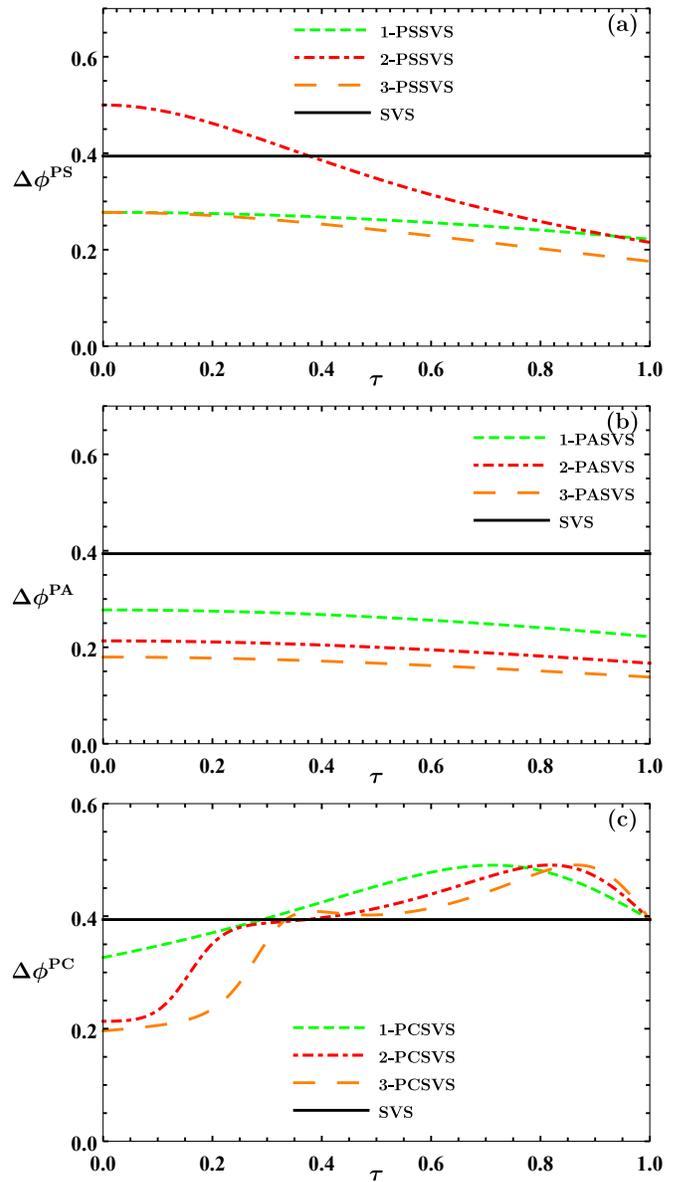}
		\caption{Phase uncertainty $\Delta \phi $ as a function of the transmissivity of the beam splitter $\tau$ for NGSVSs. We have set the squeezing parameter $r=0.5$ and the phase to be $\phi=0.01$ for all the panels.}
		\label{phase_1d_t}
	\end{center}
\end{figure}

We now study the dependence of $\Delta \phi$ on the transmissivity while other parameters are kept fixed. The results are shown in Fig.~\ref{phase_1d_t}. While phase sensitivity is maximized in the unit transmissivity limit for PSSVSs and PASVSs, phase sensitivity is maximized in the zero transmissivity limit for PCSVSs. While for 1-PSSVS, the phase sensitivity is enhanced beyond a threshold transmissivity, 2-PSSVS and 3-PSSVS improve phase sensitivity for the entire range of transmissivity. We obtain improved phase sensitivity for PASVSs for the entire range of transmissivity. In contrast, PCSVSs show improved phase sensitivity small range of low transmissivity.

\begin{figure}
	\begin{center}
		\includegraphics[scale=1]{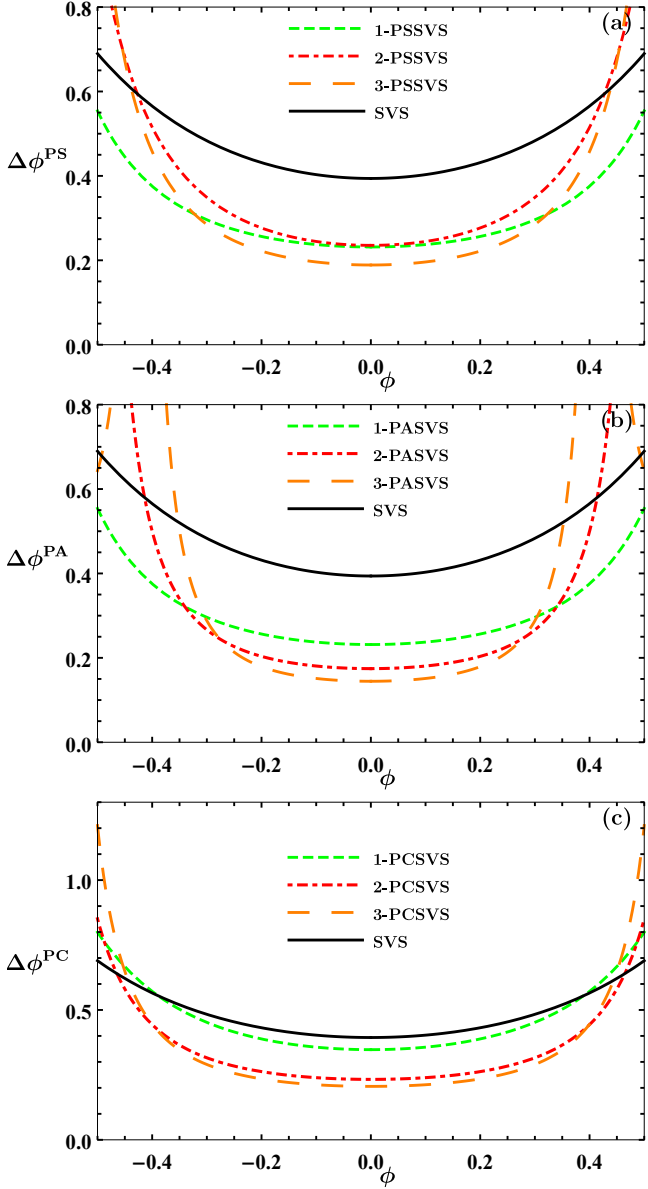}
		\caption{Phase uncertainty $\Delta \phi $ as a function of the phase $\phi$ for NGSVSs. We have set the transmissivity of the beam splitter to be $\tau=0.9$ for panels (a) and (b) and $\tau=0.1$  for panel (c). Further, the squeezing parameter has been set to be $r=0.5$ for all the panels.}
		\label{phase_1d_p}
	\end{center}
\end{figure}
In Fig.~\ref{phase_1d_p}, we analyze the dependence of $\Delta \phi$ on the phase while other parameters are kept fixed.  We notice a general trend that performing multiple NG operations results in the enhancement of phase sensitivity. However, a deviation is observed, where 1-PS operation performs better than 2-PS operation.

	\subsection{Optimal NG operation for phase estimation}
	In the preceding section, we investigated the benefits of performing NG operations on SVS  for specific values of state parameters ($r$ and $\tau$)  and phase  $\phi$. We now analyze the benefits of performing NG operations for a range of squeezing and transmissivity parameters at a fixed phase.
	This study enables us to get a good understanding of the effects of the NG operations.
	To this end, we consider the difference of $\Delta\phi$ between SVS and NGSVSs  defined as follows:
	\begin{equation}\label{merit}
\mathcal{D} ^{\text{NG} }= \Delta\phi^\text{SVS}- \Delta\phi^\text{NGSVSs}.
\end{equation}
The region of state parameters ($r$ and $\tau$), where $\mathcal{D} ^{\text{NG} }$  turns out to be positive, signifies that NGSVSs yield better phase sensitivity than the SVS.

	We also consider the success probability of the NG operations and plot them alongside the $\mathcal{D}^{\text{NG} }$ plots. Success probability signifies the fraction of successful NG operations and represents resource utilization. A careful comparison with the $\mathcal{D} ^{\text{NG} }$ plots enables us to qualitatively identify the optimal NG operation.

	\begin{figure}[h!]
		\begin{center}
			\includegraphics[scale=1]{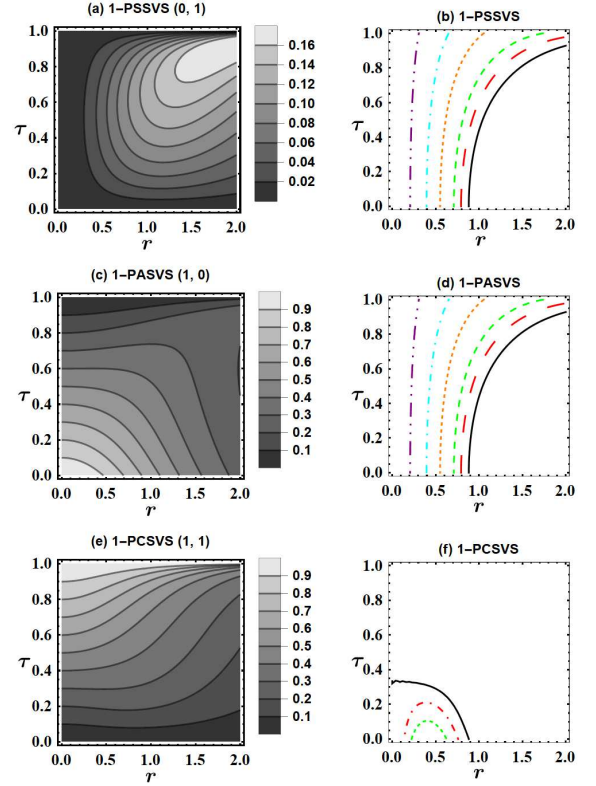}
			\caption{   Left panels depicts the success probability as a function of the transmissivity $\tau$ and
			squeezing parameter $r$ for NGSVSs. Right panels depicts   curves of  fixed $ \mathcal{D}^{\text{NG} }$, the difference of $\Delta \phi$ between SVS and NGSVSs,  as a function of $\tau$ and $r$. We have shown the  values of the parameters $(m,n)$ for different PSSVSs.   The phase, $\phi$, is taken to be $0.01$.  Solid black, large dashed red, dashed green, dotted orange, dot dashed cyan, and double dot dashed purple curves represent  $ \mathcal{D}^{\text{NG} }$~$(=0.00,\,0.025,\,0.05,\,0.10,\,0.15,\,0.20)$, respectively.}
			\label{pscontour}
		\end{center}
	\end{figure}
In the left panel of Fig.~\ref{pscontour}, we draw the contours of the success  probability in the $r$-$\tau$ space for different NG operations. We observe that the values of success probabilities reach to the range of $0.9$ for both 1-PA and 1-PC operations. However, for 1-PS operation, the success probability only reach to the range of $0.16$. For 1-PS operation, the 
 highest success probabilities are observed  for high transmissivity and high values of squeezing. In contrast, for 1-PA operation, the 
  highest success probabilities are observed  for low transmissivity and low values of squeezing. The highest success probabilities for  1-PC operation are characterized by high values of transmissivity and by low to intermediate values of squeezing in our considered range.

The right panels of Fig.~\ref{pscontour} show curves for different values of   $ \mathcal{D}^{\text{NG} }$~$(=0.00,\,0.025,\,0.05,\,0.10,\,0.15,\,0.20)$ corresponding to 1-PSSVS, 1-PASVS, and 1-PCSVS.   
For 1-PSSVS and 1-PASVS, the region of positive $ \mathcal{D}^{\text{NG}}$ is obtained for the squeezing range $r \in (0,1)$ for small values of  transmissivity. As the transmissivity increases, the advantageous squeezing range also increases. For 1-PCSVS, the region of positive $ \mathcal{D}^{\text{NG}}$ is observed for low transmissivity and low values of squeezing.

In  order to qualitatively find the most optimal NG state, we consider following two main factors: the overlap of positive regions of $ \mathcal{D}^{\text{NG} }$ with regions of high success probability, and the magnitude of the  highest success probability achived. Clearly, 1-PCSVS is out of the picture as the areas of high success probabilities do not overlap with the region with positive values of $ \mathcal{D}^{\text{NG} }$. The next scope of comparison is between 1-PSSVS and 1-PASVS where both have a considerable overlap of positive $ \mathcal{D}^{\text{NG} }$ and high success probabilities. Here, 1-PASVS turns out to be the most optimal state as the magnitude of high success probabilities ($\approx$0.9) are much greater than that of 1-PSSVS ($\approx$0.16).

	\begin{figure}
		\begin{center}
			\includegraphics[scale=1]{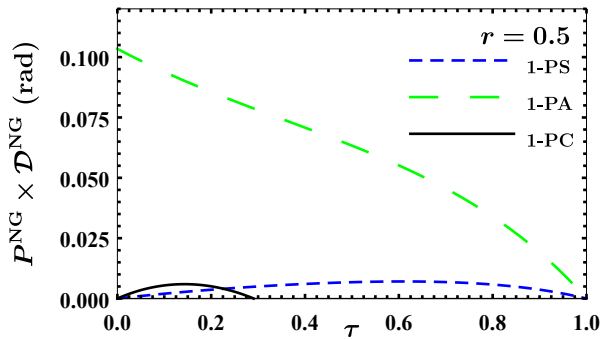}
			\caption{ Plot  of    $P^{\text{NG}} \times \mathcal{D^{\text{NG}}}  $  as a function of the transmissivity $\tau$ for different NG states. 
		  The phase has been set as $\phi=0.01$ for all the cases.}
			\label{quant}
		\end{center}
	\end{figure}

To see this comparison in a much more quantitative manner, we consider the product $ P^{\text{NG}} \times \mathcal{D^{\text{NG}}}$. Here we   trade-off
between   $P^{\text{NG}}$ and $\mathcal{D^{\text{NG}}}$  by
adjusting the transmissivity for a given squeezing to maximize the product.   The optimal state renders this product maximum. To that end, we numerically study the dependence of the product $ P^{\text{NG}} \times \mathcal{D^{\text{NG}}}$  on the transmissivity at a fixed squeezing for different NG states in Fig.~\ref{quant}. The results reveal that 1-PASVS  performs way better than other considered state when the success probability is taken into consideration.

	\section{Conclusion}\label{sec:conc}
	In this paper, we investigated the advantages offered by non-Gaussian operations in phase estimation using a parity-detection-based MZI, with coherent state and  NGSVSs as the two inputs. We considered the realistic scheme for implementing three different NG operations, namely, PS, PA, and PC, on the SVS state. We derived the Wigner function for the three corresponding NGSVSs, \ie,  PSSVSs, PASVSs, and PCSVSs. Wigner function is then used to derive the phase sensitivity of parity-detection-based MZI. The investigation of the phase sensitivity reveals that all the three NG operations can enhance the phase sensitivity for suitable choices of parameters. Further, we have taken   the success probability of different NG operations into account.
	
	The results show that the optimal operation for phase estimation is single photon addition on SVS. This is because the parameters range of high success probability for single PA operation and the large enhancement in the phase sensitivity by 1-PASVS coincide~[Fig.~\ref{pscontour}(b)].   We would like to stress that our scheme for NG state generation can be realized with currently available technologies and, therefore  is of direct relevance to the experimental community. In contrast, Refs.~\cite{gerry14,photonadded2019} has considered  photon annihilation and creation operator for the implementation of PS and PA operations, which are nonphysical. In addition, our considered figure of merit also enables us to find optimal squeezing and transmissivity parameters.
	
	Our study can be extended in several directions.
	Lang and Caves have reported that for an interferometer with a coherent state being one input and the other being constrained by average photon number, the optimal state to inject through the second input is squeezed vacuum state (SVS)~\cite{caves-prl-2013}. In a similar spirit, Ref.~\cite{photonadded2019} has compared the   phase sensitivity of ideal PSSVSs and PASVSs with  a constraint on the average photon number. It would be interesting to compare  the   phase sensitivity of NG states, including PCSVSs generated by a realistic scheme under such constraints. Further, we can also explore different measurements-based MZI, such as intensity measurement~\cite{ataman} and homodyne measurement~\cite{homodyne}. Furthermore, such an analysis involving realistic NG operation schemes can be extended to different classes of states, such as displaced Fock states~\cite{Anirban}.

	\section*{Acknowledgement}
	This is the fourth article in a publication series written in the celebration of the completion of 15 years of IISER Mohali.
	We thank   Raman Choudhary for reading the final version of the manuscript.
	C.K.  acknowledges the financial
	support from {\bf DST/ICPS/QuST/Theme-1/2019/General} Project
	number {\sf Q-68}.

	\appendix
	
	\section{Calculation of Wigner distribution function for NGSVSs}\label{app:wigner}

	\begin{figure}[h!]
		\includegraphics[scale=1]{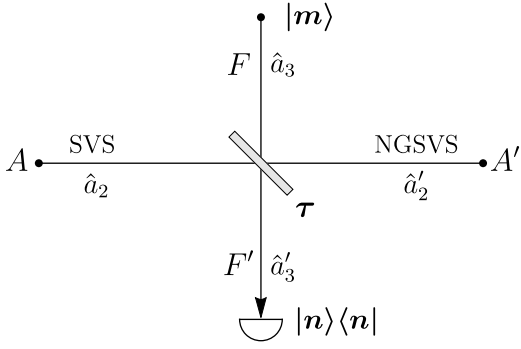}
		\caption{Schematic representation of photon subtraction, addition and catalysis operations on SVS. A beam splitter of transmissivity $\tau$ is used to mix the  SVS and the ancilla Fock state $|m\rangle$. Detection of $n$ photons  in the ancilla output mode $F^{'}$ heralds the generation of the NGSVSs.  }
		\label{figsub}
	\end{figure}
	
	In this Appendix, we provide a detailed and step wise calculation of the Wigner distribution function for the NGSVSs. The scheme for the generation of the NGSVSs is illustrated in Fig.~\ref{figsub}. We start with a single mode SVS which can be written as
 \begin{equation}
     |\text{SVS}\rangle = \mathcal{U}(S(r)) |0\rangle,
 \end{equation}
where $\mathcal{U}(S(r)) = 
\exp[r(\hat{a}_2^2-\hat{a}_2{^{\dagger}}^2)/2]$ is the single mode squeezing operator.
 This is a Gaussian state with zero mean and the following covariance matrix:
\begin{equation}
V=\frac{1}{2}\left(\begin{array}{cc}
e^{-2 r} & 0 \\
0 & e^{2 r}
\end{array}\right).
\end{equation}
The Wigner distribution function for the SVS turns out to be~\cite{weedbrook-rmp-2012}
\begin{equation}
    W(\xi_2)= \pi^{-1} \exp(-e^{-2r}q_2^2-e^{2r}p_2^2),
\end{equation}
where $\xi_2=(q_2,p_2)^T$. As shown in Fig.~\ref{figsub}, the SVS in mode A is combined with the Fock state $|m\rangle$ in the ancilla mode F using a beam-splitter  of transmissivity $\tau$. The state of the two mode system before the beam splitter transformation can be represented by its Wigner distribution function as follows:  
	\begin{equation}
		W_{  A  F}(\xi) =  W_{A}(\xi_2 ) W_{|m\rangle}(\xi_3),
	\end{equation}
		where the Wigner distribution function of a Fock state $|m\rangle$ is given by
	\begin{equation}\label{wig:fock}
		W_{|m\rangle}(q,p)=\frac{(-1)^m}{\pi}\exp  \left( 
		-q^2-p^2 \right)\,L_{m}\left[ 2(q^2+p^2) \right],
	\end{equation}
	with $L_m\{\bullet\}$ being the Laguerre polynomial of nth order.
	The action of the   beam-splitter    operation on the phase space variables $( \xi_2,\xi_3 )^T$ is given by the symplectic matrix
	\begin{equation}\label{beamsplitter}
		B_{A F}(\tau) = \begin{pmatrix}
			\sqrt{\tau} \,\mathbb{1}_2& \sqrt{1-\tau} \,\mathbb{1}_2 \\
			-\sqrt{1-\tau} \,\mathbb{1}_2&\sqrt{\tau} \,\mathbb{1}_2
		\end{pmatrix}.
	\end{equation}
	The beam splitter entangles the two modes and the corresponding  Wigner distribution function of the entangled state can be written as 
	\begin{equation}
		\begin{aligned}
			W_{  A'  F'}(\xi)   =W_{A  F}(    B_{A F}(\tau)^{-1}\xi) .
		\end{aligned}
	\end{equation}
	We now perform a conditional measurement on the ancilla mode of the output state $F'$ using a 
	 photon-number-resolving detector.
	 Detection of $n$ photons corresponds to successful implementation of NG operation on the SVS. The unnormalized Wigner distribution function of the NGSVSs  will be 
		\begin{equation}\label{detect}
		\begin{aligned}
			\widetilde{W}^{\text{NG}}_{A'  }(\xi_2 )= 2 \pi\int  d^2 \xi_3  
		 W_{A'  F'}(\xi_2,\xi_3  ) 
			\times 
			\underbrace{W_{|n \rangle }(\xi_3)}_{\text{Projection on }|n\rangle \langle n|} .\\
		\end{aligned}
	\end{equation}
	   The cases $m <n$ and $m >n$  correspond to the implementation of PS and PA operations on the SVS, respectively, while  $m=n$ corresponds to the implementation of PC operation on the SVS.  
	  PS and PA operations on SVS produce  PSSVSs and PASVSs, respectively. Similarly, PC   operation on SVS produces   PCSVSs. The states generated by performing these NG operations are NG.
	The following identity for the Laguerre polynomials can be used to transform the integrand of Eq.~(\ref{detect}) into a Gaussian function:
	\begin{equation}\label{gen}
		\begin{aligned}
			L_n[2(q^2+p^2)]=	\bm{\widehat{D}}\exp \left[\frac{st}{2}+s(q+ip)-t(q-ip)\right],
		\end{aligned}
	\end{equation}
	where the differential operator $\bm{\widehat{D}}$ is given by
	\begin{equation}
		\bm{\widehat{D}} = \frac{2^n}{n!}  \frac{\partial^n}{\partial\,s^n} \frac{\partial^n}{\partial\,t^n} \{ \bullet \}_{s=t=0}.
	\end{equation}
	   The transformed expression~(\ref{detect}) can be readily integrated  to obtain
	 \begin{equation}\label{eq4}
	\widetilde{W}^{\text{NG}}_{A'}(\xi_2 ) = \frac{\bm{\widehat{F}_1}}{\sqrt{w_1 w_2}} \exp \left(\frac{w_1^2 q_2 + w_2^2 p_2 + \bm{u}^T M_1 \bm{u}+ \bm{u}^T M_2}{-w_1 w_2}\right),
\end{equation}
where $w_{1,2} = \cosh{r} \pm \tau \sinh{r}$, column vector $\bm{u}$ is defined as $\bm{u}=(u_1,v_1,u_2,v_2)^T$, 
and differential operator $\bm{\widehat{F}_1} $ is defined as 
\begin{equation}
	\bm{\widehat{F}_1} = \frac{(-2)^{m+n}}{\pi \, m! \, n!} \frac{\partial^{m}}{\partial\,u_1^{m}} \frac{\partial^{m}}{\partial\,v_1^{m}} \frac{\partial^{n}}{\partial\,u_2^{n}} \frac{\partial^{n}}{\partial\,v_2^{n}}\{ \bullet \}_{\substack{u_1= v_1=0\\ u_2= v_2=0}}.\\
\end{equation}
Further, the   matrix $M_1$ is given by
	\begin{equation}\label{wm1}
		M_1 = \frac{1}{4}
	\left(
\begin{array}{cccc}
 \alpha  \beta  t'^2 t^2 & -\beta ^2 t'^2 & \alpha  \beta  t'^2 t & \alpha ^2 t'^2 t+t \\
 -\beta ^2 t'^2 & \alpha  \beta  t'^2 t^2 & \alpha ^2 t'^2 t+t & \alpha  \beta  t'^2 t \\
 \alpha  \beta  t'^2 t & \alpha ^2 t'^2 t+t & \alpha  \beta  t'^2 & -\alpha ^2 t'^2 t^2 \\
 \alpha ^2 t'^2 t+t & \alpha  \beta  t'^2 t & -\alpha ^2 t'^2 t^2 & \alpha  \beta  t'^2 \\
\end{array}
\right),
	\end{equation}
	where $t=\sqrt{\tau}$, $t'=\sqrt{1-\tau}$, $\alpha=\sinh \, r$ and $\beta=\cosh \, r$. The matrix $M_2$ is given by
	\begin{equation}\label{wm2}
		M_2 =
		\left(
\begin{array}{c}
- \beta  t' (q_2 w_1+i p_2 w_2) \\
 \beta  t' (q_2 w_1-i p_2 w_2) \\
- \alpha  t' t (q_2 w_1-i p_2 w_2) \\
 \alpha  t' t (q_2 w_1+i p_2 w_2) \\
\end{array}
\right).
	\end{equation}
    The probability of successful generation of NG states can be evaluated by integrating the unnormalized Wigner distribution function of the NGSVSs~(\ref{eq4}):
    \begin{equation}\label{probeqq}
		P^{\text{NG}}= \int d^2 \xi_2  \widetilde{W}^{\text{NG}}_{A' }(\xi_2)
		=\frac{ \pi \, \bm{\widehat{F}_1}}{\sqrt{w_1 w_2}} \exp \left(\frac{\bm{u}^T M_3 \bm{u}}{-4 w_1 w_2}\right),
	\end{equation}
where the matrix $M_3$ is represented as below:	
\begin{equation}\label{wm3}
		M_3 =
	\left(
\begin{array}{cccc}
 \alpha  \beta  t'^2 t^2 & \beta ^2 t'^2 & \alpha  \beta t'^2 t w_0^2 & t+\alpha ^2 t'^2 t\\
 \beta ^2 t'^2 & \alpha  \beta  t'^2 t^2 & t+\alpha ^2 t'^2 t & \alpha  \beta t'^2 t w_0^2 \\
 \alpha  \beta t'^2 t w_0^2 & t+\alpha ^2 t'^2 t & \alpha  \beta  t'^2 & \alpha ^2 t'^2 t^2 \\
 t+\alpha ^2 t'^2 t & \alpha  \beta t'^2 t w_0^2 & \alpha ^2 t'^2 t^2 & \alpha  \beta  t'^2 \\
\end{array}
\right),
	\end{equation}
	where $w_0=e^{-2r}(w_2+t'^2\alpha^2)/(w_1-t'^2\alpha^2)$. The normalized  Wigner distribution function $W^{\text{NG}}_{A'}$ of the NGSVSs can be written as follows:
	\begin{equation}\label{app:normPS}
		W^{\text{NG}}_{A' }(\xi_2 ) ={\left(P^{\text{NG}}\right)}^{-1}\widetilde{W}^{\text{NG}}_{A'}(\xi_2 ).
	\end{equation}

\section{  Matrices appearing in the average of parity operator}\label{appsec}
 Here we provide the expressions of the matrices $M_4$, $M_5$, and $M_6$  which appear in the average of parity operator~(\ref{apari}):
 \begin{equation}\label{wm4}
		M_4 = \left(
\begin{array}{cccc}
 \alpha  \beta  \gamma ^2 t'^2 t^2 & -\beta ^2 \gamma  t'^2 & \alpha  \beta  \gamma t'^2 t  & t w_3 w_4 \\
 -\beta ^2 \gamma  t'^2 & \alpha  \beta  \gamma ^2 t'^2 t^2 & t w_3 w_4 & \alpha  \beta  \gamma t'^2 t \\
 \alpha  \beta  \gamma t'^2 t & t w_3 w_4 & \alpha  \beta  t'^2 & -\alpha ^2 \gamma  t'^2 t^2 \\
 t w_3 w_4 & \alpha  \beta  \gamma t'^2 t & -\alpha ^2 \gamma t'^2 t^2 & \alpha  \beta  t'^2 \\
\end{array}
\right),
	\end{equation}
where $\gamma=\cos{\phi}$ and $\delta=\sin{\phi}$. Further,	
\begin{equation}\label{wm5}
		M_5 = \left(
\begin{array}{cc}
 \beta  \delta t' w_3  & i \beta  \delta t' w_4 \\
 -\beta  \delta t' w_3 & i \beta  \delta t' w_4 \\
 \alpha  \delta t' t w_3 & -i \alpha  \delta t' t w_4 \\
 -\alpha  \delta t' t w_3 & -i \alpha  \delta t' t w_4 \\
\end{array}
\right),
	\end{equation}
and
\begin{equation}\label{wm6}
		M_6 =
	\sin ^2\left(\frac{\phi }{2}\right) \left(
\begin{array}{cc}
 w_3 w_1  & 0 \\
 0 & w_4 w_2  \\
\end{array}
\right).
	\end{equation}
 

\begin{thebibliography}{49}%
		\makeatletter
		\providecommand \@ifxundefined [1]{%
			\@ifx{#1\undefined}
		}%
		\providecommand \@ifnum [1]{%
			\ifnum #1\expandafter \@firstoftwo
			\else \expandafter \@secondoftwo
			\fi
		}%
		\providecommand \@ifx [1]{%
			\ifx #1\expandafter \@firstoftwo
			\else \expandafter \@secondoftwo
			\fi
		}%
		\providecommand \natexlab [1]{#1}%
		\providecommand \enquote  [1]{``#1''}%
		\providecommand \bibnamefont  [1]{#1}%
		\providecommand \bibfnamefont [1]{#1}%
		\providecommand \citenamefont [1]{#1}%
		\providecommand \href@noop [0]{\@secondoftwo}%
		\providecommand \href [0]{\begingroup \@sanitize@url \@href}%
		\providecommand \@href[1]{\@@startlink{#1}\@@href}%
		\providecommand \@@href[1]{\endgroup#1\@@endlink}%
		\providecommand \@sanitize@url [0]{\catcode `\\12\catcode `\$12\catcode
			`\&12\catcode `\#12\catcode `\^12\catcode `\_12\catcode `\%12\relax}%
		\providecommand \@@startlink[1]{}%
		\providecommand \@@endlink[0]{}%
		\providecommand \url  [0]{\begingroup\@sanitize@url \@url }%
		\providecommand \@url [1]{\endgroup\@href {#1}{\urlprefix }}%
		\providecommand \urlprefix  [0]{URL }%
		\providecommand \Eprint [0]{\href }%
		\providecommand \doibase [0]{https://doi.org/}%
		\providecommand \selectlanguage [0]{\@gobble}%
		\providecommand \bibinfo  [0]{\@secondoftwo}%
		\providecommand \bibfield  [0]{\@secondoftwo}%
		\providecommand \translation [1]{[#1]}%
		\providecommand \BibitemOpen [0]{}%
		\providecommand \bibitemStop [0]{}%
		\providecommand \bibitemNoStop [0]{.\EOS\space}%
		\providecommand \EOS [0]{\spacefactor3000\relax}%
		\providecommand \BibitemShut  [1]{\csname bibitem#1\endcsname}%
		\let\auto@bib@innerbib\@empty
		\bibitem [{\citenamefont {Dowling}(2008)}]{Dowling-cp-2008}%
		\BibitemOpen
		\bibfield  {author} {\bibinfo {author} {\bibfnamefont {J.~P.}\ \bibnamefont
				{Dowling}},\ }\bibfield  {title} {\bibinfo {title} {Quantum optical metrology
				– the lowdown on high-n00n states},\ }\href
		{https://doi.org/10.1080/00107510802091298} {\bibfield  {journal} {\bibinfo
				{journal} {Contemporary Physics}\ }\textbf {\bibinfo {volume} {49}},\
			\bibinfo {pages} {125} (\bibinfo {year} {2008})}\BibitemShut {NoStop}%
		\bibitem [{\citenamefont {Giovannetti}\ \emph {et~al.}(2011)\citenamefont
			{Giovannetti}, \citenamefont {Lloyd},\ and\ \citenamefont
			{Maccone}}]{Giovannetti2011}%
		\BibitemOpen
		\bibfield  {author} {\bibinfo {author} {\bibfnamefont {V.}~\bibnamefont
				{Giovannetti}}, \bibinfo {author} {\bibfnamefont {S.}~\bibnamefont {Lloyd}},\
			and\ \bibinfo {author} {\bibfnamefont {L.}~\bibnamefont {Maccone}},\
		}\bibfield  {title} {\bibinfo {title} {Advances in quantum metrology},\
		}\href {https://doi.org/10.1038/nphoton.2011.35} {\bibfield  {journal}
			{\bibinfo  {journal} {Nature Photonics}\ }\textbf {\bibinfo {volume} {5}},\
			\bibinfo {pages} {222} (\bibinfo {year} {2011})}\BibitemShut {NoStop}%
		\bibitem [{\citenamefont {Caves}(1981)}]{caves-prd-1981}%
		\BibitemOpen
		\bibfield  {author} {\bibinfo {author} {\bibfnamefont {C.~M.}\ \bibnamefont
				{Caves}},\ }\bibfield  {title} {\bibinfo {title} {Quantum-mechanical noise in
				an interferometer},\ }\href {https://doi.org/10.1103/PhysRevD.23.1693}
		{\bibfield  {journal} {\bibinfo  {journal} {Phys. Rev. D}\ }\textbf {\bibinfo
				{volume} {23}},\ \bibinfo {pages} {1693} (\bibinfo {year}
			{1981})}\BibitemShut {NoStop}%
		\bibitem [{\citenamefont {Boto}\ \emph {et~al.}(2000)\citenamefont {Boto},
			\citenamefont {Kok}, \citenamefont {Abrams}, \citenamefont {Braunstein},
			\citenamefont {Williams},\ and\ \citenamefont {Dowling}}]{Dowling}%
		\BibitemOpen
		\bibfield  {author} {\bibinfo {author} {\bibfnamefont {A.~N.}\ \bibnamefont
				{Boto}}, \bibinfo {author} {\bibfnamefont {P.}~\bibnamefont {Kok}}, \bibinfo
			{author} {\bibfnamefont {D.~S.}\ \bibnamefont {Abrams}}, \bibinfo {author}
			{\bibfnamefont {S.~L.}\ \bibnamefont {Braunstein}}, \bibinfo {author}
			{\bibfnamefont {C.~P.}\ \bibnamefont {Williams}},\ and\ \bibinfo {author}
			{\bibfnamefont {J.~P.}\ \bibnamefont {Dowling}},\ }\bibfield  {title}
		{\bibinfo {title} {Quantum interferometric optical lithography: Exploiting
				entanglement to beat the diffraction limit},\ }\href
		{https://doi.org/10.1103/PhysRevLett.85.2733} {\bibfield  {journal} {\bibinfo
				{journal} {Phys. Rev. Lett.}\ }\textbf {\bibinfo {volume} {85}},\ \bibinfo
			{pages} {2733} (\bibinfo {year} {2000})}\BibitemShut {NoStop}%
		\bibitem [{\citenamefont {Giovannetti}\ \emph {et~al.}(2004)\citenamefont
			{Giovannetti}, \citenamefont {Lloyd},\ and\ \citenamefont
			{Maccone}}]{Giovannetti-science-2004}%
		\BibitemOpen
		\bibfield  {author} {\bibinfo {author} {\bibfnamefont {V.}~\bibnamefont
				{Giovannetti}}, \bibinfo {author} {\bibfnamefont {S.}~\bibnamefont {Lloyd}},\
			and\ \bibinfo {author} {\bibfnamefont {L.}~\bibnamefont {Maccone}},\
		}\bibfield  {title} {\bibinfo {title} {Quantum-enhanced measurements: Beating
				the standard quantum limit},\ }\href
		{https://doi.org/10.1126/science.1104149} {\bibfield  {journal} {\bibinfo
				{journal} {Science}\ }\textbf {\bibinfo {volume} {306}},\ \bibinfo {pages}
			{1330} (\bibinfo {year} {2004})}\BibitemShut {NoStop}%
		\bibitem [{\citenamefont {Hofmann}\ and\ \citenamefont
			{Ono}(2007)}]{Hofmann-pra-2007}%
		\BibitemOpen
		\bibfield  {author} {\bibinfo {author} {\bibfnamefont {H.~F.}\ \bibnamefont
				{Hofmann}}\ and\ \bibinfo {author} {\bibfnamefont {T.}~\bibnamefont {Ono}},\
		}\bibfield  {title} {\bibinfo {title} {High-photon-number path entanglement
				in the interference of spontaneously down-converted photon pairs with
				coherent laser light},\ }\href {https://doi.org/10.1103/PhysRevA.76.031806}
		{\bibfield  {journal} {\bibinfo  {journal} {Phys. Rev. A}\ }\textbf {\bibinfo
				{volume} {76}},\ \bibinfo {pages} {031806} (\bibinfo {year}
			{2007})}\BibitemShut {NoStop}%
		\bibitem [{\citenamefont {Anisimov}\ \emph {et~al.}(2010)\citenamefont
			{Anisimov}, \citenamefont {Raterman}, \citenamefont {Chiruvelli},
			\citenamefont {Plick}, \citenamefont {Huver}, \citenamefont {Lee},\ and\
			\citenamefont {Dowling}}]{Anisimov-prl-2010}%
		\BibitemOpen
		\bibfield  {author} {\bibinfo {author} {\bibfnamefont {P.~M.}\ \bibnamefont
				{Anisimov}}, \bibinfo {author} {\bibfnamefont {G.~M.}\ \bibnamefont
				{Raterman}}, \bibinfo {author} {\bibfnamefont {A.}~\bibnamefont
				{Chiruvelli}}, \bibinfo {author} {\bibfnamefont {W.~N.}\ \bibnamefont
				{Plick}}, \bibinfo {author} {\bibfnamefont {S.~D.}\ \bibnamefont {Huver}},
			\bibinfo {author} {\bibfnamefont {H.}~\bibnamefont {Lee}},\ and\ \bibinfo
			{author} {\bibfnamefont {J.~P.}\ \bibnamefont {Dowling}},\ }\bibfield
		{title} {\bibinfo {title} {Quantum metrology with two-mode squeezed vacuum:
				Parity detection beats the heisenberg limit},\ }\href
		{https://doi.org/10.1103/PhysRevLett.104.103602} {\bibfield  {journal}
			{\bibinfo  {journal} {Phys. Rev. Lett.}\ }\textbf {\bibinfo {volume} {104}},\
			\bibinfo {pages} {103602} (\bibinfo {year} {2010})}\BibitemShut {NoStop}%
		\bibitem [{\citenamefont {Lang}\ and\ \citenamefont
			{Caves}(2013)}]{caves-prl-2013}%
		\BibitemOpen
		\bibfield  {author} {\bibinfo {author} {\bibfnamefont {M.~D.}\ \bibnamefont
				{Lang}}\ and\ \bibinfo {author} {\bibfnamefont {C.~M.}\ \bibnamefont
				{Caves}},\ }\bibfield  {title} {\bibinfo {title} {Optimal quantum-enhanced
				interferometry using a laser power source},\ }\href
		{https://doi.org/10.1103/PhysRevLett.111.173601} {\bibfield  {journal}
			{\bibinfo  {journal} {Phys. Rev. Lett.}\ }\textbf {\bibinfo {volume} {111}},\
			\bibinfo {pages} {173601} (\bibinfo {year} {2013})}\BibitemShut {NoStop}%
		\bibitem [{\citenamefont {Kwon}\ \emph {et~al.}(2019)\citenamefont {Kwon},
			\citenamefont {Tan}, \citenamefont {Volkoff},\ and\ \citenamefont
			{Jeong}}]{Jeong-prl-2019}%
		\BibitemOpen
		\bibfield  {author} {\bibinfo {author} {\bibfnamefont {H.}~\bibnamefont
				{Kwon}}, \bibinfo {author} {\bibfnamefont {K.~C.}\ \bibnamefont {Tan}},
			\bibinfo {author} {\bibfnamefont {T.}~\bibnamefont {Volkoff}},\ and\ \bibinfo
			{author} {\bibfnamefont {H.}~\bibnamefont {Jeong}},\ }\bibfield  {title}
		{\bibinfo {title} {Nonclassicality as a quantifiable resource for quantum
				metrology},\ }\href {https://doi.org/10.1103/PhysRevLett.122.040503}
		{\bibfield  {journal} {\bibinfo  {journal} {Phys. Rev. Lett.}\ }\textbf
			{\bibinfo {volume} {122}},\ \bibinfo {pages} {040503} (\bibinfo {year}
			{2019})}\BibitemShut {NoStop}%
		\bibitem [{\citenamefont {Vahlbruch}\ \emph {et~al.}(2016)\citenamefont
			{Vahlbruch}, \citenamefont {Mehmet}, \citenamefont {Danzmann},\ and\
			\citenamefont {Schnabel}}]{15dB}%
		\BibitemOpen
		\bibfield  {author} {\bibinfo {author} {\bibfnamefont {H.}~\bibnamefont
				{Vahlbruch}}, \bibinfo {author} {\bibfnamefont {M.}~\bibnamefont {Mehmet}},
			\bibinfo {author} {\bibfnamefont {K.}~\bibnamefont {Danzmann}},\ and\
			\bibinfo {author} {\bibfnamefont {R.}~\bibnamefont {Schnabel}},\ }\bibfield
		{title} {\bibinfo {title} {Detection of 15 db squeezed states of light and
				their application for the absolute calibration of photoelectric quantum
				efficiency},\ }\href {https://doi.org/10.1103/PhysRevLett.117.110801}
		{\bibfield  {journal} {\bibinfo  {journal} {Phys. Rev. Lett.}\ }\textbf
			{\bibinfo {volume} {117}},\ \bibinfo {pages} {110801} (\bibinfo {year}
			{2016})}\BibitemShut {NoStop}%
		\bibitem [{\citenamefont {Opatrn\'y}\ \emph {et~al.}(2000)\citenamefont
			{Opatrn\'y}, \citenamefont {Kurizki},\ and\ \citenamefont
			{Welsch}}]{tel2000}%
		\BibitemOpen
		\bibfield  {author} {\bibinfo {author} {\bibfnamefont {T.}~\bibnamefont
				{Opatrn\'y}}, \bibinfo {author} {\bibfnamefont {G.}~\bibnamefont {Kurizki}},\
			and\ \bibinfo {author} {\bibfnamefont {D.-G.}\ \bibnamefont {Welsch}},\
		}\bibfield  {title} {\bibinfo {title} {Improvement on teleportation of
				continuous variables by photon subtraction via conditional measurement},\
		}\href {https://doi.org/10.1103/PhysRevA.61.032302} {\bibfield  {journal}
			{\bibinfo  {journal} {Phys. Rev. A}\ }\textbf {\bibinfo {volume} {61}},\
			\bibinfo {pages} {032302} (\bibinfo {year} {2000})}\BibitemShut {NoStop}%
		\bibitem [{\citenamefont {Kitagawa}\ \emph {et~al.}(2006)\citenamefont
			{Kitagawa}, \citenamefont {Takeoka}, \citenamefont {Sasaki},\ and\
			\citenamefont {Chefles}}]{Akira-pra-2006}%
		\BibitemOpen
		\bibfield  {author} {\bibinfo {author} {\bibfnamefont {A.}~\bibnamefont
				{Kitagawa}}, \bibinfo {author} {\bibfnamefont {M.}~\bibnamefont {Takeoka}},
			\bibinfo {author} {\bibfnamefont {M.}~\bibnamefont {Sasaki}},\ and\ \bibinfo
			{author} {\bibfnamefont {A.}~\bibnamefont {Chefles}},\ }\bibfield  {title}
		{\bibinfo {title} {Entanglement evaluation of non-gaussian states generated
				by photon subtraction from squeezed states},\ }\href
		{https://doi.org/10.1103/PhysRevA.73.042310} {\bibfield  {journal} {\bibinfo
				{journal} {Phys. Rev. A}\ }\textbf {\bibinfo {volume} {73}},\ \bibinfo
			{pages} {042310} (\bibinfo {year} {2006})}\BibitemShut {NoStop}%
		\bibitem [{\citenamefont {Dell'Anno}\ \emph {et~al.}(2007)\citenamefont
			{Dell'Anno}, \citenamefont {De~Siena}, \citenamefont {Albano},\ and\
			\citenamefont {Illuminati}}]{Anno-2007}%
		\BibitemOpen
		\bibfield  {author} {\bibinfo {author} {\bibfnamefont {F.}~\bibnamefont
				{Dell'Anno}}, \bibinfo {author} {\bibfnamefont {S.}~\bibnamefont {De~Siena}},
			\bibinfo {author} {\bibfnamefont {L.}~\bibnamefont {Albano}},\ and\ \bibinfo
			{author} {\bibfnamefont {F.}~\bibnamefont {Illuminati}},\ }\bibfield  {title}
		{\bibinfo {title} {Continuous-variable quantum teleportation with
				non-gaussian resources},\ }\href {https://doi.org/10.1103/PhysRevA.76.022301}
		{\bibfield  {journal} {\bibinfo  {journal} {Phys. Rev. A}\ }\textbf {\bibinfo
				{volume} {76}},\ \bibinfo {pages} {022301} (\bibinfo {year}
			{2007})}\BibitemShut {NoStop}%
		\bibitem [{\citenamefont {Yang}\ and\ \citenamefont {Li}(2009)}]{tel2009}%
		\BibitemOpen
		\bibfield  {author} {\bibinfo {author} {\bibfnamefont {Y.}~\bibnamefont
				{Yang}}\ and\ \bibinfo {author} {\bibfnamefont {F.-L.}\ \bibnamefont {Li}},\
		}\bibfield  {title} {\bibinfo {title} {Entanglement properties of
				non-gaussian resources generated via photon subtraction and addition and
				continuous-variable quantum-teleportation improvement},\ }\href
		{https://doi.org/10.1103/PhysRevA.80.022315} {\bibfield  {journal} {\bibinfo
				{journal} {Phys. Rev. A}\ }\textbf {\bibinfo {volume} {80}},\ \bibinfo
			{pages} {022315} (\bibinfo {year} {2009})}\BibitemShut {NoStop}%
		\bibitem [{\citenamefont {Wang}\ \emph {et~al.}(2015)\citenamefont {Wang},
			\citenamefont {Hou}, \citenamefont {Chen},\ and\ \citenamefont
			{Xu}}]{wang2015}%
		\BibitemOpen
		\bibfield  {author} {\bibinfo {author} {\bibfnamefont {S.}~\bibnamefont
				{Wang}}, \bibinfo {author} {\bibfnamefont {L.-L.}\ \bibnamefont {Hou}},
			\bibinfo {author} {\bibfnamefont {X.-F.}\ \bibnamefont {Chen}},\ and\
			\bibinfo {author} {\bibfnamefont {X.-F.}\ \bibnamefont {Xu}},\ }\bibfield
		{title} {\bibinfo {title} {Continuous-variable quantum teleportation with
				non-gaussian entangled states generated via multiple-photon subtraction and
				addition},\ }\href {https://doi.org/10.1103/PhysRevA.91.063832} {\bibfield
			{journal} {\bibinfo  {journal} {Phys. Rev. A}\ }\textbf {\bibinfo {volume}
				{91}},\ \bibinfo {pages} {063832} (\bibinfo {year} {2015})}\BibitemShut
		{NoStop}%
		\bibitem [{\citenamefont {Xu}(2015)}]{catalysis15}%
		\BibitemOpen
		\bibfield  {author} {\bibinfo {author} {\bibfnamefont {X.-x.}\ \bibnamefont
				{Xu}},\ }\bibfield  {title} {\bibinfo {title} {Enhancing quantum entanglement
				and quantum teleportation for two-mode squeezed vacuum state by local
				quantum-optical catalysis},\ }\href
		{https://doi.org/10.1103/PhysRevA.92.012318} {\bibfield  {journal} {\bibinfo
				{journal} {Phys. Rev. A}\ }\textbf {\bibinfo {volume} {92}},\ \bibinfo
			{pages} {012318} (\bibinfo {year} {2015})}\BibitemShut {NoStop}%
		\bibitem [{\citenamefont {Hu}\ \emph {et~al.}(2017)\citenamefont {Hu},
			\citenamefont {Liao},\ and\ \citenamefont {Zubairy}}]{catalysis17}%
		\BibitemOpen
		\bibfield  {author} {\bibinfo {author} {\bibfnamefont {L.}~\bibnamefont
				{Hu}}, \bibinfo {author} {\bibfnamefont {Z.}~\bibnamefont {Liao}},\ and\
			\bibinfo {author} {\bibfnamefont {M.~S.}\ \bibnamefont {Zubairy}},\
		}\bibfield  {title} {\bibinfo {title} {Continuous-variable entanglement via
				multiphoton catalysis},\ }\href {https://doi.org/10.1103/PhysRevA.95.012310}
		{\bibfield  {journal} {\bibinfo  {journal} {Phys. Rev. A}\ }\textbf {\bibinfo
				{volume} {95}},\ \bibinfo {pages} {012310} (\bibinfo {year}
			{2017})}\BibitemShut {NoStop}%
		\bibitem [{\citenamefont {Kumar}\ and\ \citenamefont
			{Arora}(2022)}]{tele-arxiv}%
		\BibitemOpen
		\bibfield  {author} {\bibinfo {author} {\bibfnamefont {C.}~\bibnamefont
				{Kumar}}\ and\ \bibinfo {author} {\bibfnamefont {S.}~\bibnamefont {Arora}},\
		}\bibfield  {title} {\bibinfo {title} {Experimental-schemes-based
				non-gaussian operations in continuous variable quantum teleportation},\
		}\href {https://doi.org/10.48550/arXiv.2206.06806} {\bibfield  {journal}
			{\bibinfo  {journal} {arxiv.2206.06806}\ } (\bibinfo {year}
			{2022})}\BibitemShut {NoStop}%
		\bibitem [{\citenamefont {Huang}\ \emph {et~al.}(2013)\citenamefont {Huang},
			\citenamefont {He}, \citenamefont {Fang},\ and\ \citenamefont
			{Zeng}}]{qkd-pra-2013}%
		\BibitemOpen
		\bibfield  {author} {\bibinfo {author} {\bibfnamefont {P.}~\bibnamefont
				{Huang}}, \bibinfo {author} {\bibfnamefont {G.}~\bibnamefont {He}}, \bibinfo
			{author} {\bibfnamefont {J.}~\bibnamefont {Fang}},\ and\ \bibinfo {author}
			{\bibfnamefont {G.}~\bibnamefont {Zeng}},\ }\bibfield  {title} {\bibinfo
			{title} {Performance improvement of continuous-variable quantum key
				distribution via photon subtraction},\ }\href
		{https://doi.org/10.1103/PhysRevA.87.012317} {\bibfield  {journal} {\bibinfo
				{journal} {Phys. Rev. A}\ }\textbf {\bibinfo {volume} {87}},\ \bibinfo
			{pages} {012317} (\bibinfo {year} {2013})}\BibitemShut {NoStop}%
		\bibitem [{\citenamefont {Ma}\ \emph {et~al.}(2018)\citenamefont {Ma},
			\citenamefont {Huang}, \citenamefont {Bai}, \citenamefont {Wang},
			\citenamefont {Bao},\ and\ \citenamefont {Zeng}}]{qkd-pra-2018}%
		\BibitemOpen
		\bibfield  {author} {\bibinfo {author} {\bibfnamefont {H.-X.}\ \bibnamefont
				{Ma}}, \bibinfo {author} {\bibfnamefont {P.}~\bibnamefont {Huang}}, \bibinfo
			{author} {\bibfnamefont {D.-Y.}\ \bibnamefont {Bai}}, \bibinfo {author}
			{\bibfnamefont {S.-Y.}\ \bibnamefont {Wang}}, \bibinfo {author}
			{\bibfnamefont {W.-S.}\ \bibnamefont {Bao}},\ and\ \bibinfo {author}
			{\bibfnamefont {G.-H.}\ \bibnamefont {Zeng}},\ }\bibfield  {title} {\bibinfo
			{title} {Continuous-variable measurement-device-independent quantum key
				distribution with photon subtraction},\ }\href
		{https://doi.org/10.1103/PhysRevA.97.042329} {\bibfield  {journal} {\bibinfo
				{journal} {Phys. Rev. A}\ }\textbf {\bibinfo {volume} {97}},\ \bibinfo
			{pages} {042329} (\bibinfo {year} {2018})}\BibitemShut {NoStop}%
		\bibitem [{\citenamefont {Guo}\ \emph {et~al.}(2019)\citenamefont {Guo},
			\citenamefont {Ye}, \citenamefont {Zhong},\ and\ \citenamefont
			{Liao}}]{qkd-pra-2019}%
		\BibitemOpen
		\bibfield  {author} {\bibinfo {author} {\bibfnamefont {Y.}~\bibnamefont
				{Guo}}, \bibinfo {author} {\bibfnamefont {W.}~\bibnamefont {Ye}}, \bibinfo
			{author} {\bibfnamefont {H.}~\bibnamefont {Zhong}},\ and\ \bibinfo {author}
			{\bibfnamefont {Q.}~\bibnamefont {Liao}},\ }\bibfield  {title} {\bibinfo
			{title} {Continuous-variable quantum key distribution with non-gaussian
				quantum catalysis},\ }\href {https://doi.org/10.1103/PhysRevA.99.032327}
		{\bibfield  {journal} {\bibinfo  {journal} {Phys. Rev. A}\ }\textbf {\bibinfo
				{volume} {99}},\ \bibinfo {pages} {032327} (\bibinfo {year}
			{2019})}\BibitemShut {NoStop}%
		\bibitem [{\citenamefont {Ye}\ \emph {et~al.}(2019)\citenamefont {Ye},
			\citenamefont {Zhong}, \citenamefont {Liao}, \citenamefont {Huang},
			\citenamefont {Hu},\ and\ \citenamefont {Guo}}]{qk2019}%
		\BibitemOpen
		\bibfield  {author} {\bibinfo {author} {\bibfnamefont {W.}~\bibnamefont
				{Ye}}, \bibinfo {author} {\bibfnamefont {H.}~\bibnamefont {Zhong}}, \bibinfo
			{author} {\bibfnamefont {Q.}~\bibnamefont {Liao}}, \bibinfo {author}
			{\bibfnamefont {D.}~\bibnamefont {Huang}}, \bibinfo {author} {\bibfnamefont
				{L.}~\bibnamefont {Hu}},\ and\ \bibinfo {author} {\bibfnamefont
				{Y.}~\bibnamefont {Guo}},\ }\bibfield  {title} {\bibinfo {title} {Improvement
				of self-referenced continuous-variable quantum key distribution with quantum
				photon catalysis},\ }\href {https://doi.org/10.1364/OE.27.017186} {\bibfield
			{journal} {\bibinfo  {journal} {Opt. Express}\ }\textbf {\bibinfo {volume}
				{27}},\ \bibinfo {pages} {17186} (\bibinfo {year} {2019})}\BibitemShut
		{NoStop}%
		\bibitem [{\citenamefont {Hu}\ \emph {et~al.}(2020)\citenamefont {Hu},
			\citenamefont {Al-amri}, \citenamefont {Liao},\ and\ \citenamefont
			{Zubairy}}]{zubairy-pra-2020}%
		\BibitemOpen
		\bibfield  {author} {\bibinfo {author} {\bibfnamefont {L.}~\bibnamefont
				{Hu}}, \bibinfo {author} {\bibfnamefont {M.}~\bibnamefont {Al-amri}},
			\bibinfo {author} {\bibfnamefont {Z.}~\bibnamefont {Liao}},\ and\ \bibinfo
			{author} {\bibfnamefont {M.~S.}\ \bibnamefont {Zubairy}},\ }\bibfield
		{title} {\bibinfo {title} {Continuous-variable quantum key distribution with
				non-gaussian operations},\ }\href
		{https://doi.org/10.1103/PhysRevA.102.012608} {\bibfield  {journal} {\bibinfo
				{journal} {Phys. Rev. A}\ }\textbf {\bibinfo {volume} {102}},\ \bibinfo
			{pages} {012608} (\bibinfo {year} {2020})}\BibitemShut {NoStop}%
		\bibitem [{\citenamefont {Zhang}\ \emph {et~al.}(2014)\citenamefont {Zhang},
			\citenamefont {Guo}, \citenamefont {Bao}, \citenamefont {Shi}, \citenamefont
			{Jin}, \citenamefont {Zou},\ and\ \citenamefont {Guo}}]{illumination14}%
		\BibitemOpen
		\bibfield  {author} {\bibinfo {author} {\bibfnamefont {S.}~\bibnamefont
				{Zhang}}, \bibinfo {author} {\bibfnamefont {J.}~\bibnamefont {Guo}}, \bibinfo
			{author} {\bibfnamefont {W.}~\bibnamefont {Bao}}, \bibinfo {author}
			{\bibfnamefont {J.}~\bibnamefont {Shi}}, \bibinfo {author} {\bibfnamefont
				{C.}~\bibnamefont {Jin}}, \bibinfo {author} {\bibfnamefont {X.}~\bibnamefont
				{Zou}},\ and\ \bibinfo {author} {\bibfnamefont {G.}~\bibnamefont {Guo}},\
		}\bibfield  {title} {\bibinfo {title} {Quantum illumination with
				photon-subtracted continuous-variable entanglement},\ }\href
		{https://doi.org/10.1103/PhysRevA.89.062309} {\bibfield  {journal} {\bibinfo
				{journal} {Phys. Rev. A}\ }\textbf {\bibinfo {volume} {89}},\ \bibinfo
			{pages} {062309} (\bibinfo {year} {2014})}\BibitemShut {NoStop}%
		\bibitem [{\citenamefont {Birrittella}\ \emph {et~al.}(2012)\citenamefont
			{Birrittella}, \citenamefont {Mimih},\ and\ \citenamefont
			{Gerry}}]{gerryc-pra-2012}%
		\BibitemOpen
		\bibfield  {author} {\bibinfo {author} {\bibfnamefont {R.}~\bibnamefont
				{Birrittella}}, \bibinfo {author} {\bibfnamefont {J.}~\bibnamefont {Mimih}},\
			and\ \bibinfo {author} {\bibfnamefont {C.~C.}\ \bibnamefont {Gerry}},\
		}\bibfield  {title} {\bibinfo {title} {Multiphoton quantum interference at a
				beam splitter and the approach to heisenberg-limited interferometry},\ }\href
		{https://doi.org/10.1103/PhysRevA.86.063828} {\bibfield  {journal} {\bibinfo
				{journal} {Phys. Rev. A}\ }\textbf {\bibinfo {volume} {86}},\ \bibinfo
			{pages} {063828} (\bibinfo {year} {2012})}\BibitemShut {NoStop}%
		\bibitem [{\citenamefont {Carranza}\ and\ \citenamefont
			{Gerry}(2012)}]{josab-2012}%
		\BibitemOpen
		\bibfield  {author} {\bibinfo {author} {\bibfnamefont {R.}~\bibnamefont
				{Carranza}}\ and\ \bibinfo {author} {\bibfnamefont {C.~C.}\ \bibnamefont
				{Gerry}},\ }\bibfield  {title} {\bibinfo {title} {Photon-subtracted two-mode
				squeezed vacuum states and applications to quantum optical interferometry},\
		}\href {https://doi.org/10.1364/JOSAB.29.002581} {\bibfield  {journal}
			{\bibinfo  {journal} {J. Opt. Soc. Am. B}\ }\textbf {\bibinfo {volume}
				{29}},\ \bibinfo {pages} {2581} (\bibinfo {year} {2012})}\BibitemShut
		{NoStop}%
		\bibitem [{\citenamefont {Braun}\ \emph {et~al.}(2014)\citenamefont {Braun},
			\citenamefont {Jian}, \citenamefont {Pinel},\ and\ \citenamefont
			{Treps}}]{braun-pra-2014}%
		\BibitemOpen
		\bibfield  {author} {\bibinfo {author} {\bibfnamefont {D.}~\bibnamefont
				{Braun}}, \bibinfo {author} {\bibfnamefont {P.}~\bibnamefont {Jian}},
			\bibinfo {author} {\bibfnamefont {O.}~\bibnamefont {Pinel}},\ and\ \bibinfo
			{author} {\bibfnamefont {N.}~\bibnamefont {Treps}},\ }\bibfield  {title}
		{\bibinfo {title} {Precision measurements with photon-subtracted or
				photon-added gaussian states},\ }\href
		{https://doi.org/10.1103/PhysRevA.90.013821} {\bibfield  {journal} {\bibinfo
				{journal} {Phys. Rev. A}\ }\textbf {\bibinfo {volume} {90}},\ \bibinfo
			{pages} {013821} (\bibinfo {year} {2014})}\BibitemShut {NoStop}%
		\bibitem [{\citenamefont {Ouyang}\ \emph {et~al.}(2016)\citenamefont {Ouyang},
			\citenamefont {Wang},\ and\ \citenamefont {Zhang}}]{josab-2016}%
		\BibitemOpen
		\bibfield  {author} {\bibinfo {author} {\bibfnamefont {Y.}~\bibnamefont
				{Ouyang}}, \bibinfo {author} {\bibfnamefont {S.}~\bibnamefont {Wang}},\ and\
			\bibinfo {author} {\bibfnamefont {L.}~\bibnamefont {Zhang}},\ }\bibfield
		{title} {\bibinfo {title} {Quantum optical interferometry via the
				photon-added two-mode squeezed vacuum states},\ }\href
		{https://doi.org/10.1364/JOSAB.33.001373} {\bibfield  {journal} {\bibinfo
				{journal} {J. Opt. Soc. Am. B}\ }\textbf {\bibinfo {volume} {33}},\ \bibinfo
			{pages} {1373} (\bibinfo {year} {2016})}\BibitemShut {NoStop}%
		\bibitem [{\citenamefont {Zhang}\ \emph {et~al.}(2021)\citenamefont {Zhang},
			\citenamefont {Ye}, \citenamefont {Wei}, \citenamefont {Xia}, \citenamefont
			{Chang}, \citenamefont {Liao},\ and\ \citenamefont
			{Hu}}]{pra-catalysis-2021}%
		\BibitemOpen
		\bibfield  {author} {\bibinfo {author} {\bibfnamefont {H.}~\bibnamefont
				{Zhang}}, \bibinfo {author} {\bibfnamefont {W.}~\bibnamefont {Ye}}, \bibinfo
			{author} {\bibfnamefont {C.}~\bibnamefont {Wei}}, \bibinfo {author}
			{\bibfnamefont {Y.}~\bibnamefont {Xia}}, \bibinfo {author} {\bibfnamefont
				{S.}~\bibnamefont {Chang}}, \bibinfo {author} {\bibfnamefont
				{Z.}~\bibnamefont {Liao}},\ and\ \bibinfo {author} {\bibfnamefont
				{L.}~\bibnamefont {Hu}},\ }\bibfield  {title} {\bibinfo {title} {Improved
				phase sensitivity in a quantum optical interferometer based on multiphoton
				catalytic two-mode squeezed vacuum states},\ }\href
		{https://doi.org/10.1103/PhysRevA.103.013705} {\bibfield  {journal} {\bibinfo
				{journal} {Phys. Rev. A}\ }\textbf {\bibinfo {volume} {103}},\ \bibinfo
			{pages} {013705} (\bibinfo {year} {2021})}\BibitemShut {NoStop}%
		\bibitem [{\citenamefont {Kumar}\ \emph
			{et~al.}(2022{\natexlab{a}})\citenamefont {Kumar}, \citenamefont {Rishabh},\
			and\ \citenamefont {Arora}}]{crs-ngtmsv-met}%
		\BibitemOpen
		\bibfield  {author} {\bibinfo {author} {\bibfnamefont {C.}~\bibnamefont
				{Kumar}}, \bibinfo {author} {\bibnamefont {Rishabh}},\ and\ \bibinfo {author}
			{\bibfnamefont {S.}~\bibnamefont {Arora}},\ }\bibfield  {title} {\bibinfo
			{title} {Realistic non-gaussian-operation scheme in parity-detection-based
				mach-zehnder quantum interferometry},\ }\href
		{https://doi.org/10.1103/PhysRevA.105.052437} {\bibfield  {journal} {\bibinfo
				{journal} {Phys. Rev. A}\ }\textbf {\bibinfo {volume} {105}},\ \bibinfo
			{pages} {052437} (\bibinfo {year} {2022}{\natexlab{a}})}\BibitemShut
		{NoStop}%
		\bibitem [{\citenamefont {Kumar}\ \emph
			{et~al.}(2022{\natexlab{b}})\citenamefont {Kumar}, \citenamefont
			{{Rishabh}},\ and\ \citenamefont {Arora}}]{metro-thermal-arxiv}%
		\BibitemOpen
		\bibfield  {author} {\bibinfo {author} {\bibfnamefont {C.}~\bibnamefont
				{Kumar}}, \bibinfo {author} {\bibnamefont {{Rishabh}}},\ and\ \bibinfo
			{author} {\bibfnamefont {S.}~\bibnamefont {Arora}},\ }\bibfield  {title}
		{\bibinfo {title} {Enhanced phase estimation in parity detection based
				mach-zehnder interferometer using non-gaussian two-mode squeezed thermal
				input state},\ }\href {https://doi.org/10.48550/arXiv.2208.04742} {\bibfield
			{journal} {\bibinfo  {journal} {arxiv.2208.04742}\ } (\bibinfo {year}
			{2022}{\natexlab{b}})}\BibitemShut {NoStop}%
		\bibitem [{\citenamefont {Birrittella}\ and\ \citenamefont
			{Gerry}(2014)}]{gerry14}%
		\BibitemOpen
		\bibfield  {author} {\bibinfo {author} {\bibfnamefont {R.}~\bibnamefont
				{Birrittella}}\ and\ \bibinfo {author} {\bibfnamefont {C.~C.}\ \bibnamefont
				{Gerry}},\ }\bibfield  {title} {\bibinfo {title} {Quantum optical
				interferometry via the mixing of coherent and photon-subtracted squeezed
				vacuum states of light},\ }\href {https://doi.org/10.1364/JOSAB.31.000586}
		{\bibfield  {journal} {\bibinfo  {journal} {J. Opt. Soc. Am. B}\ }\textbf
			{\bibinfo {volume} {31}},\ \bibinfo {pages} {586} (\bibinfo {year}
			{2014})}\BibitemShut {NoStop}%
		\bibitem [{\citenamefont {Lvovsky}\ \emph {et~al.}(2001)\citenamefont
			{Lvovsky}, \citenamefont {Hansen}, \citenamefont {Aichele}, \citenamefont
			{Benson}, \citenamefont {Mlynek},\ and\ \citenamefont
			{Schiller}}]{singlephoton}%
		\BibitemOpen
		\bibfield  {author} {\bibinfo {author} {\bibfnamefont {A.~I.}\ \bibnamefont
				{Lvovsky}}, \bibinfo {author} {\bibfnamefont {H.}~\bibnamefont {Hansen}},
			\bibinfo {author} {\bibfnamefont {T.}~\bibnamefont {Aichele}}, \bibinfo
			{author} {\bibfnamefont {O.}~\bibnamefont {Benson}}, \bibinfo {author}
			{\bibfnamefont {J.}~\bibnamefont {Mlynek}},\ and\ \bibinfo {author}
			{\bibfnamefont {S.}~\bibnamefont {Schiller}},\ }\bibfield  {title} {\bibinfo
			{title} {Quantum state reconstruction of the single-photon fock state},\
		}\href {https://doi.org/10.1103/PhysRevLett.87.050402} {\bibfield  {journal}
			{\bibinfo  {journal} {Phys. Rev. Lett.}\ }\textbf {\bibinfo {volume} {87}},\
			\bibinfo {pages} {050402} (\bibinfo {year} {2001})}\BibitemShut {NoStop}%
		\bibitem [{\citenamefont {Zavatta}\ \emph {et~al.}(2004)\citenamefont
			{Zavatta}, \citenamefont {Viciani},\ and\ \citenamefont
			{Bellini}}]{single-photon}%
		\BibitemOpen
		\bibfield  {author} {\bibinfo {author} {\bibfnamefont {A.}~\bibnamefont
				{Zavatta}}, \bibinfo {author} {\bibfnamefont {S.}~\bibnamefont {Viciani}},\
			and\ \bibinfo {author} {\bibfnamefont {M.}~\bibnamefont {Bellini}},\
		}\bibfield  {title} {\bibinfo {title} {Tomographic reconstruction of the
				single-photon fock state by high-frequency homodyne detection},\ }\href
		{https://doi.org/10.1103/PhysRevA.70.053821} {\bibfield  {journal} {\bibinfo
				{journal} {Phys. Rev. A}\ }\textbf {\bibinfo {volume} {70}},\ \bibinfo
			{pages} {053821} (\bibinfo {year} {2004})}\BibitemShut {NoStop}%
		\bibitem [{\citenamefont {Huisman}\ \emph {et~al.}(2009)\citenamefont
			{Huisman}, \citenamefont {Jain}, \citenamefont {Babichev}, \citenamefont
			{Vewinger}, \citenamefont {Zhang}, \citenamefont {Youn},\ and\ \citenamefont
			{Lvovsky}}]{singlephot}%
		\BibitemOpen
		\bibfield  {author} {\bibinfo {author} {\bibfnamefont {S.~R.}\ \bibnamefont
				{Huisman}}, \bibinfo {author} {\bibfnamefont {N.}~\bibnamefont {Jain}},
			\bibinfo {author} {\bibfnamefont {S.~A.}\ \bibnamefont {Babichev}}, \bibinfo
			{author} {\bibfnamefont {F.}~\bibnamefont {Vewinger}}, \bibinfo {author}
			{\bibfnamefont {A.~N.}\ \bibnamefont {Zhang}}, \bibinfo {author}
			{\bibfnamefont {S.~H.}\ \bibnamefont {Youn}},\ and\ \bibinfo {author}
			{\bibfnamefont {A.~I.}\ \bibnamefont {Lvovsky}},\ }\bibfield  {title}
		{\bibinfo {title} {Instant single-photon fock state tomography},\ }\href
		{https://doi.org/10.1364/OL.34.002739} {\bibfield  {journal} {\bibinfo
				{journal} {Opt. Lett.}\ }\textbf {\bibinfo {volume} {34}},\ \bibinfo {pages}
			{2739} (\bibinfo {year} {2009})}\BibitemShut {NoStop}%
		\bibitem [{\citenamefont {Ourjoumtsev}\ \emph {et~al.}(2006)\citenamefont
			{Ourjoumtsev}, \citenamefont {Tualle-Brouri},\ and\ \citenamefont
			{Grangier}}]{2phton}%
		\BibitemOpen
		\bibfield  {author} {\bibinfo {author} {\bibfnamefont {A.}~\bibnamefont
				{Ourjoumtsev}}, \bibinfo {author} {\bibfnamefont {R.}~\bibnamefont
				{Tualle-Brouri}},\ and\ \bibinfo {author} {\bibfnamefont {P.}~\bibnamefont
				{Grangier}},\ }\bibfield  {title} {\bibinfo {title} {Quantum homodyne
				tomography of a two-photon fock state},\ }\href
		{https://doi.org/10.1103/PhysRevLett.96.213601} {\bibfield  {journal}
			{\bibinfo  {journal} {Phys. Rev. Lett.}\ }\textbf {\bibinfo {volume} {96}},\
			\bibinfo {pages} {213601} (\bibinfo {year} {2006})}\BibitemShut {NoStop}%
		\bibitem [{\citenamefont {Cooper}\ \emph {et~al.}(2013)\citenamefont {Cooper},
			\citenamefont {Wright}, \citenamefont {S\"{o}ller},\ and\ \citenamefont
			{Smith}}]{3phton}%
		\BibitemOpen
		\bibfield  {author} {\bibinfo {author} {\bibfnamefont {M.}~\bibnamefont
				{Cooper}}, \bibinfo {author} {\bibfnamefont {L.~J.}\ \bibnamefont {Wright}},
			\bibinfo {author} {\bibfnamefont {C.}~\bibnamefont {S\"{o}ller}},\ and\
			\bibinfo {author} {\bibfnamefont {B.~J.}\ \bibnamefont {Smith}},\ }\bibfield
		{title} {\bibinfo {title} {Experimental generation of multi-photon fock
				states},\ }\href {https://doi.org/10.1364/OE.21.005309} {\bibfield  {journal}
			{\bibinfo  {journal} {Opt. Express}\ }\textbf {\bibinfo {volume} {21}},\
			\bibinfo {pages} {5309} (\bibinfo {year} {2013})}\BibitemShut {NoStop}%
		\bibitem [{\citenamefont {Lita}\ \emph {et~al.}(2008)\citenamefont {Lita},
			\citenamefont {Miller},\ and\ \citenamefont {Nam}}]{Lita:08}%
		\BibitemOpen
		\bibfield  {author} {\bibinfo {author} {\bibfnamefont {A.~E.}\ \bibnamefont
				{Lita}}, \bibinfo {author} {\bibfnamefont {A.~J.}\ \bibnamefont {Miller}},\
			and\ \bibinfo {author} {\bibfnamefont {S.~W.}\ \bibnamefont {Nam}},\
		}\bibfield  {title} {\bibinfo {title} {Counting near-infrared single-photons
				with 95\% efficiency},\ }\href {https://doi.org/10.1364/OE.16.003032}
		{\bibfield  {journal} {\bibinfo  {journal} {Opt. Express}\ }\textbf {\bibinfo
				{volume} {16}},\ \bibinfo {pages} {3032} (\bibinfo {year}
			{2008})}\BibitemShut {NoStop}%
		\bibitem [{\citenamefont {Marsili}\ \emph {et~al.}(2013)\citenamefont
			{Marsili}, \citenamefont {Verma}, \citenamefont {Stern}, \citenamefont
			{Harrington}, \citenamefont {Lita}, \citenamefont {Gerrits}, \citenamefont
			{Vayshenker}, \citenamefont {Baek}, \citenamefont {Shaw}, \citenamefont
			{Mirin},\ and\ \citenamefont {Nam}}]{Marsili2013}%
		\BibitemOpen
		\bibfield  {author} {\bibinfo {author} {\bibfnamefont {F.}~\bibnamefont
				{Marsili}}, \bibinfo {author} {\bibfnamefont {V.~B.}\ \bibnamefont {Verma}},
			\bibinfo {author} {\bibfnamefont {J.~A.}\ \bibnamefont {Stern}}, \bibinfo
			{author} {\bibfnamefont {S.}~\bibnamefont {Harrington}}, \bibinfo {author}
			{\bibfnamefont {A.~E.}\ \bibnamefont {Lita}}, \bibinfo {author}
			{\bibfnamefont {T.}~\bibnamefont {Gerrits}}, \bibinfo {author} {\bibfnamefont
				{I.}~\bibnamefont {Vayshenker}}, \bibinfo {author} {\bibfnamefont
				{B.}~\bibnamefont {Baek}}, \bibinfo {author} {\bibfnamefont {M.~D.}\
				\bibnamefont {Shaw}}, \bibinfo {author} {\bibfnamefont {R.~P.}\ \bibnamefont
				{Mirin}},\ and\ \bibinfo {author} {\bibfnamefont {S.~W.}\ \bibnamefont
				{Nam}},\ }\bibfield  {title} {\bibinfo {title} {Detecting single infrared
				photons with 93{\%} system efficiency},\ }\href
		{https://doi.org/10.1038/nphoton.2013.13} {\bibfield  {journal} {\bibinfo
				{journal} {Nature Photonics}\ }\textbf {\bibinfo {volume} {7}},\ \bibinfo
			{pages} {210} (\bibinfo {year} {2013})}\BibitemShut {NoStop}%
		\bibitem [{\citenamefont {Esmaeil~Zadeh}\ \emph {et~al.}(2017)\citenamefont
			{Esmaeil~Zadeh}, \citenamefont {Los}, \citenamefont {Gourgues}, \citenamefont
			{Steinmetz}, \citenamefont {Bulgarini}, \citenamefont {Dobrovolskiy},
			\citenamefont {Zwiller},\ and\ \citenamefont {Dorenbos}}]{Zadeh}%
		\BibitemOpen
		\bibfield  {author} {\bibinfo {author} {\bibfnamefont {I.}~\bibnamefont
				{Esmaeil~Zadeh}}, \bibinfo {author} {\bibfnamefont {J.~W.~N.}\ \bibnamefont
				{Los}}, \bibinfo {author} {\bibfnamefont {R.~B.~M.}\ \bibnamefont
				{Gourgues}}, \bibinfo {author} {\bibfnamefont {V.}~\bibnamefont {Steinmetz}},
			\bibinfo {author} {\bibfnamefont {G.}~\bibnamefont {Bulgarini}}, \bibinfo
			{author} {\bibfnamefont {S.~M.}\ \bibnamefont {Dobrovolskiy}}, \bibinfo
			{author} {\bibfnamefont {V.}~\bibnamefont {Zwiller}},\ and\ \bibinfo {author}
			{\bibfnamefont {S.~N.}\ \bibnamefont {Dorenbos}},\ }\bibfield  {title}
		{\bibinfo {title} {Single-photon detectors combining high efficiency, high
				detection rates, and ultra-high timing resolution},\ }\href
		{https://doi.org/10.1063/1.5000001} {\bibfield  {journal} {\bibinfo
				{journal} {APL Photonics}\ }\textbf {\bibinfo {volume} {2}},\ \bibinfo
			{pages} {111301} (\bibinfo {year} {2017})}\BibitemShut {NoStop}%
		\bibitem [{\citenamefont {Wang}\ \emph {et~al.}(2019)\citenamefont {Wang},
			\citenamefont {Xu}, \citenamefont {Xu},\ and\ \citenamefont
			{Zhang}}]{photonadded2019}%
		\BibitemOpen
		\bibfield  {author} {\bibinfo {author} {\bibfnamefont {S.}~\bibnamefont
				{Wang}}, \bibinfo {author} {\bibfnamefont {X.}~\bibnamefont {Xu}}, \bibinfo
			{author} {\bibfnamefont {Y.}~\bibnamefont {Xu}},\ and\ \bibinfo {author}
			{\bibfnamefont {L.}~\bibnamefont {Zhang}},\ }\bibfield  {title} {\bibinfo
			{title} {Quantum interferometry via a coherent state mixed with a
				photon-added squeezed vacuum state},\ }\href
		{https://doi.org/https://doi.org/10.1016/j.optcom.2019.03.068} {\bibfield
			{journal} {\bibinfo  {journal} {Optics Communications}\ }\textbf {\bibinfo
				{volume} {444}},\ \bibinfo {pages} {102} (\bibinfo {year}
			{2019})}\BibitemShut {NoStop}%
		\bibitem [{\citenamefont {Yurke}\ \emph {et~al.}(1986)\citenamefont {Yurke},
			\citenamefont {McCall},\ and\ \citenamefont {Klauder}}]{yurke-1986}%
		\BibitemOpen
		\bibfield  {author} {\bibinfo {author} {\bibfnamefont {B.}~\bibnamefont
				{Yurke}}, \bibinfo {author} {\bibfnamefont {S.~L.}\ \bibnamefont {McCall}},\
			and\ \bibinfo {author} {\bibfnamefont {J.~R.}\ \bibnamefont {Klauder}},\
		}\bibfield  {title} {\bibinfo {title} {Su(2) and su(1,1) interferometers},\
		}\href {https://doi.org/10.1103/PhysRevA.33.4033} {\bibfield  {journal}
			{\bibinfo  {journal} {Phys. Rev. A}\ }\textbf {\bibinfo {volume} {33}},\
			\bibinfo {pages} {4033} (\bibinfo {year} {1986})}\BibitemShut {NoStop}%
		\bibitem [{\citenamefont {Arvind}\ \emph {et~al.}(1995)\citenamefont {Arvind},
			\citenamefont {Dutta}, \citenamefont {Mukunda},\ and\ \citenamefont
			{Simon}}]{arvind1995}%
		\BibitemOpen
		\bibfield  {author} {\bibinfo {author} {\bibnamefont {Arvind}}, \bibinfo
			{author} {\bibfnamefont {B.}~\bibnamefont {Dutta}}, \bibinfo {author}
			{\bibfnamefont {N.}~\bibnamefont {Mukunda}},\ and\ \bibinfo {author}
			{\bibfnamefont {R.}~\bibnamefont {Simon}},\ }\bibfield  {title} {\bibinfo
			{title} {The real symplectic groups in quantum mechanics and optics},\ }\href
		{https://doi.org/10.1007/BF02848172} {\bibfield  {journal} {\bibinfo
				{journal} {Pramana}\ }\textbf {\bibinfo {volume} {45}},\ \bibinfo {pages}
			{471} (\bibinfo {year} {1995})}\BibitemShut {NoStop}%
		\bibitem [{\citenamefont {Weedbrook}\ \emph {et~al.}(2012)\citenamefont
			{Weedbrook}, \citenamefont {Pirandola}, \citenamefont {Garc\'{\i}a-Patr\'on},
			\citenamefont {Cerf}, \citenamefont {Ralph}, \citenamefont {Shapiro},\ and\
			\citenamefont {Lloyd}}]{weedbrook-rmp-2012}%
		\BibitemOpen
		\bibfield  {author} {\bibinfo {author} {\bibfnamefont {C.}~\bibnamefont
				{Weedbrook}}, \bibinfo {author} {\bibfnamefont {S.}~\bibnamefont
				{Pirandola}}, \bibinfo {author} {\bibfnamefont {R.}~\bibnamefont
				{Garc\'{\i}a-Patr\'on}}, \bibinfo {author} {\bibfnamefont {N.~J.}\
				\bibnamefont {Cerf}}, \bibinfo {author} {\bibfnamefont {T.~C.}\ \bibnamefont
				{Ralph}}, \bibinfo {author} {\bibfnamefont {J.~H.}\ \bibnamefont {Shapiro}},\
			and\ \bibinfo {author} {\bibfnamefont {S.}~\bibnamefont {Lloyd}},\ }\bibfield
		{title} {\bibinfo {title} {Gaussian quantum information},\ }\href
		{https://doi.org/10.1103/RevModPhys.84.621} {\bibfield  {journal} {\bibinfo
				{journal} {Rev. Mod. Phys.}\ }\textbf {\bibinfo {volume} {84}},\ \bibinfo
			{pages} {621} (\bibinfo {year} {2012})}\BibitemShut {NoStop}%
		\bibitem [{\citenamefont {Royer}(1977)}]{parity-1977}%
		\BibitemOpen
		\bibfield  {author} {\bibinfo {author} {\bibfnamefont {A.}~\bibnamefont
				{Royer}},\ }\bibfield  {title} {\bibinfo {title} {Wigner function as the
				expectation value of a parity operator},\ }\href
		{https://doi.org/10.1103/PhysRevA.15.449} {\bibfield  {journal} {\bibinfo
				{journal} {Phys. Rev. A}\ }\textbf {\bibinfo {volume} {15}},\ \bibinfo
			{pages} {449} (\bibinfo {year} {1977})}\BibitemShut {NoStop}%
		\bibitem [{\citenamefont {Birrittella}\ \emph {et~al.}(2021)\citenamefont
			{Birrittella}, \citenamefont {Alsing},\ and\ \citenamefont
			{Gerry}}]{Birrittella-2021}%
		\BibitemOpen
		\bibfield  {author} {\bibinfo {author} {\bibfnamefont {R.~J.}\ \bibnamefont
				{Birrittella}}, \bibinfo {author} {\bibfnamefont {P.~M.}\ \bibnamefont
				{Alsing}},\ and\ \bibinfo {author} {\bibfnamefont {C.~C.}\ \bibnamefont
				{Gerry}},\ }\bibfield  {title} {\bibinfo {title} {The parity operator:
				Applications in quantum metrology},\ }\href
		{https://doi.org/10.1116/5.0026148} {\bibfield  {journal} {\bibinfo
				{journal} {AVS Quantum Science}\ }\textbf {\bibinfo {volume} {3}},\ \bibinfo
			{pages} {014701} (\bibinfo {year} {2021})}\BibitemShut {NoStop}%
		\bibitem [{\citenamefont {Ataman}\ \emph {et~al.}(2018)\citenamefont {Ataman},
			\citenamefont {Preda},\ and\ \citenamefont {Ionicioiu}}]{ataman}%
		\BibitemOpen
		\bibfield  {author} {\bibinfo {author} {\bibfnamefont {S.}~\bibnamefont
				{Ataman}}, \bibinfo {author} {\bibfnamefont {A.}~\bibnamefont {Preda}},\ and\
			\bibinfo {author} {\bibfnamefont {R.}~\bibnamefont {Ionicioiu}},\ }\bibfield
		{title} {\bibinfo {title} {Phase sensitivity of a mach-zehnder interferometer
				with single-intensity and difference-intensity detection},\ }\href
		{https://doi.org/10.1103/PhysRevA.98.043856} {\bibfield  {journal} {\bibinfo
				{journal} {Phys. Rev. A}\ }\textbf {\bibinfo {volume} {98}},\ \bibinfo
			{pages} {043856} (\bibinfo {year} {2018})}\BibitemShut {NoStop}%
		\bibitem [{\citenamefont {Bondurant}\ and\ \citenamefont
			{Shapiro}(1984)}]{homodyne}%
		\BibitemOpen
		\bibfield  {author} {\bibinfo {author} {\bibfnamefont {R.~S.}\ \bibnamefont
				{Bondurant}}\ and\ \bibinfo {author} {\bibfnamefont {J.~H.}\ \bibnamefont
				{Shapiro}},\ }\bibfield  {title} {\bibinfo {title} {Squeezed states in
				phase-sensing interferometers},\ }\href
		{https://doi.org/10.1103/PhysRevD.30.2548} {\bibfield  {journal} {\bibinfo
				{journal} {Phys. Rev. D}\ }\textbf {\bibinfo {volume} {30}},\ \bibinfo
			{pages} {2548} (\bibinfo {year} {1984})}\BibitemShut {NoStop}%
		\bibitem [{\citenamefont {Malpani}\ \emph {et~al.}(2019)\citenamefont
			{Malpani}, \citenamefont {Thapliyal}, \citenamefont {Alam}, \citenamefont
			{Pathak}, \citenamefont {Narayanan},\ and\ \citenamefont
			{Banerjee}}]{Anirban}%
		\BibitemOpen
		\bibfield  {author} {\bibinfo {author} {\bibfnamefont {P.}~\bibnamefont
				{Malpani}}, \bibinfo {author} {\bibfnamefont {K.}~\bibnamefont {Thapliyal}},
			\bibinfo {author} {\bibfnamefont {N.}~\bibnamefont {Alam}}, \bibinfo {author}
			{\bibfnamefont {A.}~\bibnamefont {Pathak}}, \bibinfo {author} {\bibfnamefont
				{V.}~\bibnamefont {Narayanan}},\ and\ \bibinfo {author} {\bibfnamefont
				{S.}~\bibnamefont {Banerjee}},\ }\bibfield  {title} {\bibinfo {title}
			{Quantum phase properties of photon added and subtracted displaced fock
				states},\ }\href {https://doi.org/https://doi.org/10.1002/andp.201900141}
		{\bibfield  {journal} {\bibinfo  {journal} {Annalen der Physik}\ }\textbf
			{\bibinfo {volume} {531}},\ \bibinfo {pages} {1900141} (\bibinfo {year}
			{2019})}\BibitemShut {NoStop}%
	\end{thebibliography}
	%
	
\end{document}